\documentclass[prb,twocolumn,amssymb,amsmath,aps,superscriptaddress,showpacs,eqsecnum]{revtex4}
\usepackage{amsfonts}
\usepackage{graphicx}
\usepackage{epsfig}
\def\bea{\begin{eqnarray}}
\def\eea{\end{eqnarray}}
\def\be{\begin{equation}}
\def\ee{\end{equation}}

\usepackage{hyperref}

\begin{document}
\title{Nonequilibrium Critical Behavior for Electron Tunneling through Quantum Dots in an Aharonov-Bohm Circuit}

\author{Eran Sela and Ian Affleck \date{\today}}
\affiliation{Department of Physics and Astronomy, University of
British Columbia, Vancouver, B.C., Canada, V6T 1Z1}

\begin{abstract}
Double quantum dots can provide an experimental realization of the 2
impurity Kondo model which exhibits a non-Fermi liquid quantum
critical point (QCP) at a special value of its parameters. We
generalize our recent study of double quantum dots in series
\cite{Sela08} to a parallel configuration with an Aharonov-Bohm
flux. We present an exact universal result for the finite
temperature and finite voltage conductance $G[V,T]$ along the
crossover from the QCP to the low energy Fermi liquid phase.
Compared to the series configuration, here generically $G[V,T] \ne
G[-V,T]$, leading to current rectification.
\end{abstract}

\maketitle

\section{Introduction}It is now well established that quantum dots (QD's) behave as Kondo
impurities at low temperatures.\cite{Gordon98,Cronenwett98} Whereas
many theoretical tools are available to address linear transport,
the nonequilibrium regime is far less studied, although it is
typically addressed in experiment.\cite{Grobis08} A solution of
nonequilibrium transport through a 1-channel Kondo impurity was
achieved in Ref.~[\onlinecite{Schiller98}]; however exact results
were obtained only for a specific point in the parameters space
(Toulouse limit). Another important development in this direction
was the application of the Bethe-ansatz and finding of many body
scattering states.\cite{Konik02,Mehta06} Recently we found exact
results for nonlinear transport close to a QCP in a double dot in
series realizing the two impurity Kondo model (2IKM).\cite{Sela08}

The 2IKM consists of two impurity spins coupled to two channels of
conduction electrons and, at the same time, interacting with each
other through an exchange interaction $K$. Jones \emph{et.
al.}~\cite{Jones88} observed that a QCP at $K=K_c$ separates a
''local singlet'' from a Kondo-screened phase, where $K_c$ is of the
order of the Kondo temperature $T_K$. The exact critical behavior
was found using conformal field theory \cite{Affleck92,Affleck95}
(CFT) and abelian bosonization~\cite{Gan95} methods. Implications of
the 2IKM for transport through double QDs were studied in
Refs.~[\onlinecite{ddot,Georges99,Izumida,Zarand06,moreddot,moreddot1,Konik}].

The presence of a sharp quantum phase transition in the 2IKM became
questionable soon after its discovery; in the mean field study in
Ref.~[\onlinecite{Jones89}] it was pointed out that the true QCP is
restricted to the case of a special particle hole (P-H) symmetry
assumed in Ref.~[\onlinecite{Jones88}]. This was confirmed by
numerical renormalization group calculations.~\cite{Sakai} P-H
symmetry breaking was later associated with two relevant potential
scattering perturbations.~\cite{Affleck95,Zarand06} Thus, in real
systems the critical behavior for $K = K_c$ can be observed only
above a certain crossover energy scale, denoted here as $T^*_{LR}$.
In order to obtain reliable predictions for QDs it is crucial to
include the extra relevant perturbations associated with potential
scattering in a real calculation. We achieved this task for a double
QD~\cite{Sela08} using the method developed by Gan.\cite{Gan95} The
finding of exact crossover results including P-H symmetry breaking
remains an open problem for the alternative proposed realization of
the 2IKM by Zar\'{a}nd \emph{et. al.}~\cite{Zarand06} Compared to
their QD system involving at least three leads, our system has only
two leads making the nonequilibrium behavior more tractable.

In this paper we generalize our previous results to a generic
configuration ranging from series to parallel QD attached to two
leads; see Fig.~(\ref{fg:1}). In this generic configuration
transport from left to right occurs via different interfering paths.
A particular feature of our results distinguishes the generic case
from the series case: in the generic case the finite voltage
conductance $G[V]$ has the property $G[V]- G[-V] \ne 0$, leading to
current rectification, similar to a diode. This effect results from
interactions and is absent in a noninteracting Landauer
description.\cite{Landauer57} An additional aim of this paper is to
provide important details on the calculation for the general series
or parallel cases.

The outline of the paper is as follows. In Sec.~\ref{se:model} the
double QD system is presented and mapped to the 2IKM. In
Sec.~\ref{se:T=0} the conductance is calculated at the QCP using CFT
methods, neglecting the effect of potentials scattering. In
Sec.~\ref{se:Tne0} we consider deviations from the QCP due to
variations of $K$ from $K_c$, and calculate the finite temperature
crossover for the linear conductance using a mapping of the P-H
symmetric 2IKM to the Ising model with a boundary magnetic field. We
also apply this mapping for the QD system proposed by Zar\'{a}nd
\emph{et. al.}~\cite{Zarand06} as a realization of the 2IKM. In
Sec.~\ref{se:HQCP} potential scattering is incorporated in the
Hamiltonian close to the QCP, and in the crossover formula for the
linear conductance. In Sec.~\ref{se:V} the full nonequilirium
problem at finite voltage and temperature in the vicinity of the
critical point is addressed. Sec.~\ref{se:conclusion} contains
conclusions. We relegate details on the calculation of the nonlinear
conductance using Keldysh Green functions (GFs) to the appendix.

\section{Model}
\label{se:model}
\begin{figure}[h]
\begin{center}
\includegraphics*[width=70mm]{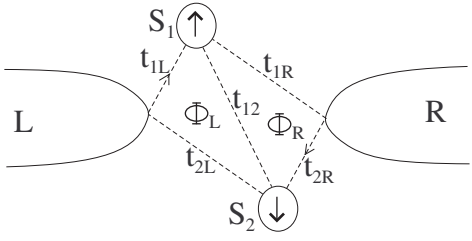}
\caption{Schematic description of the physical system, consisting of
two leads ($L$ and $R$) and two quantum dots with effective
spin-1/2. We use the convention that for finite flux $t_{1L} \to
t_{1L}e^{i \Phi_L}$ and $t_{2R} \to t_{2R}e^{i \Phi_R}$ where the
$t_{iL}$ are defined as tunneling amplitudes from lead ($L,R$) $\to$
 dot (i).\label{fg:1}}
\end{center}
\end{figure}
The physical system under consideration is shown schematically in
Fig.~(\ref{fg:1}). It consists of left $(L)$ and right $(R)$ leads
tunnel coupled to two quantum dots $1$, $2$, with tunneling
amplitudes $t_{iL/R}$, $(i=1,2)$. We assume that both dots are in
the Kondo regime, with gate voltages adjusted to give an odd number
of electrons and the $t_{iL/R}$ are sufficiently weak compared to
the charging energy, $U$, so that charge fluctuations can be
ignored. We write the effective spin-1/2 moments as $\vec{S}_1$ and
$\vec{S}_2$. We will be primarily interested in the case where
$t_{1L}$, $t_{2R}\gg t_{1R}$, $t_{2L}$ so that the left lead is
primarily coupled to dot $1$ and the right lead to dot $2$ since
only in this case will the QCP occur. Note that in the extreme case
where $t_{1R}=t_{2L}=0$, this reduces to the series configuration
analyzed, for example, in Ref.~\onlinecite{Sela08}. The fluxes
$\Phi_L$ and $\Phi_R$ are introduced in the triangular plaquettes as
shown.

In the standard fashion,~\cite{Affleck91} the conduction-electron
channels that couple to the impurity are reduced to one-dimensional
left moving Dirac fields $\psi_{i \alpha}(x)$, where $i=L,R$ and\
$\alpha=\uparrow,\downarrow$ are the lead and spin indices,
respectively. We assume that a \emph{single mode} in each lead
couples to both impurities. Here we have linearized the
conduction-electron dispersion around the Fermi level:
$\epsilon_k=\hbar v_F k$, where $\epsilon_k$ and $k$ are measured
relative to the Fermi level and Fermi wave number, respectively. $x$
is a fictitious position variable conjugate to $k$. We set $\hbar
=v_F=1$.

We discuss the different terms which will appear in the model
Hamiltonian, Eq.~(\ref{eq:H}). An exchange interaction
\begin{equation}
\label{eq:K12} K_{12}\sim \frac{t_{12}^2}{U} \end{equation} between
the impurity spins is generated by the interimpurity tunneling
$t_{12}$. The impurity spins are
also Kondo-coupled to the conduction-electron spin density at the
origin:
\begin{eqnarray}
\vec{s}^i_j = {\psi^\dagger}^{i \alpha}
\frac{\vec{\sigma}_\alpha^\beta}{2} \psi_{j
\beta},~~~i,j=L,R=1,2.\nonumber \\
(\rm{repeated~spin~indices~summed})\nonumber
\end{eqnarray}
In addition there are potential scattering (PS) terms $\propto
{\psi^\dagger}^{i \alpha}  \psi_{j \alpha}~~
(\rm{repeated~spin~indices~summed})$. The system is driven out of
equilibrium by a source drain voltage $V$. Thus, the Hamiltonian $H$
is
\begin{eqnarray}
\label{eq:H} H&=&H_0+H_V + K_{12} \vec{S}_1 \cdot \vec{S}_2+
H_{K}+H_{PS}+ H', \nonumber  \\
H_0 &=& \int_{-\infty}^\infty dx {\psi^\dagger}^{j \alpha} i
\partial_x \psi_{j \alpha}, \nonumber \\
H_V &=& \frac{eV}{2} \int_{-\infty}^\infty dx {\psi^\dagger}^{i
\alpha} (\tau^z)_i^j \psi_{j \alpha}, \nonumber \\
H_{K}&=&  \sum_{\ell = 1,2} ({J^{(\ell)}})^j_i \vec{s}^{~i}_j \cdot
\vec{S}_\ell ,\nonumber \\
H_{PS} &= & {\psi^\dagger}^{i \alpha} V_{i}^j \psi_{j \alpha},
\end{eqnarray}
with repeated lead and spin indices summed. To be complete one has
to add the terms
\begin{equation}
 H' ={{V}''}_{i}^j \vec{s}^{~i}_j \cdot
(\vec{S}_1 \times \vec{S}_2 )+{\psi^\dagger}^{i \alpha} {{V}'}_{i}^j
\psi_{j \alpha} (\vec{S}_1 \cdot \vec{S}_2 ). \nonumber
\end{equation}
However, close to the QCP the first (second) term of $H'$ has a
similar effect as $H_{K}$ ($H_{PS}$). Therefore, up to a correction
to the actual coupling constants, energy scales, and to the critical
value of different parameters at the QCP, all of which we are not
able to determine exactly, it is legitimate to drop $H'$.

The Kondo interaction induces, via the Ruderman-Kittel-Kasuya-Yosida
(RKKY) mechanism, an additional contribution to the inter-impurity
exchange, $K = K_{12}+K_{RKKY}$, where
\begin{eqnarray}
K_{RKKY} &=&2 \langle S_1 = \downarrow ,S_2 = \uparrow|H_{K}
\frac{1}{-H_0} H_{K}|S_1 = \uparrow ,S_2 = \downarrow
\rangle\nonumber \\
&=& 4 \sum_{ k_1
>0, k_2<0 } \frac{1}{-(\epsilon_{k_2} - \epsilon_{k_1})} {\rm{tr}}\{
J^{(1)} J^{(2)} \}. \nonumber
\end{eqnarray}
With the parametrization of $J^{(\ell)}$ given in
Eq.~(\ref{eq:parametrization}), ${\rm{tr}}\{ J^{(1)} J^{(2)} \} = 4
J^2 \sin^2 (2 \theta) \cos^2 \frac{\Phi}{2}$. Using $\sum_k = \nu
\int d \epsilon$, where $\nu$ is the density of states in the leads,
and restricting the band width to $|\epsilon_k| < U$, beyond which
the effective spin description breaks down, one obtains a
ferromagnetic contribution
\begin{equation} \label{eq:RKKY} K_{RKKY} \sim - ( \nu J)^2 U
\sin^2 (2 \theta)\cos^2 \frac{\Phi}{2}.
\end{equation}

We estimate the potential scattering amplitudes by
\begin{eqnarray}
\label{eq:V1}
V_i^i& \sim&\frac{t^2_{1 i}+t^2_{2 i}}{U},~~~(i=L,R) \nonumber \\
V_L^R &\sim&\frac{t_{1L} t_{1R} e^{i \Phi_L}+t_{2L} t_{2R} e^{-i
\Phi_R}}{U}\nonumber \\
&+&c' \frac{t_{1L} t_{12} t_{2R}e^{i (\Phi_L -\Phi_R )}+t_{2L}
t_{12} t_{1R}  }{U^2},
\end{eqnarray}
where $c'$ is a constant factor of order 1.

Until Sec.~\ref{se:observability} we will assume the parity symmetry
\begin{eqnarray}
\label{eq:parity}S_1 \leftrightarrow S_2, \qquad L \leftrightarrow
R.
\end{eqnarray}
However our results are not restricted to this case, as will be
discussed in Sec.~(\ref{se:observability}). Parity implies $t_{1L} =
t_{2 R} \equiv t_1$; $t_{2L} = t_{1 R} \equiv t_2$. For finite flux
$\Phi = \Phi_L+\Phi_R$ the parity symmetry is preserved for
$\Phi_L=\Phi_R$. Calculating the Kondo couplings to second order in
the tunneling amplitudes, under this symmetry, gives $\{
J^{(1)},J^{(2)} \}
\propto \{ \frac{v v^\dagger}{U} ,\frac{\tau^x v v^\dagger \tau^x}{U} \}$, where $v=\left(%
\begin{array}{cc}
 t_1 e^{i \Phi/2} \\
  t_2 \\
\end{array}%
\right)$. This leads us to parameterize the hermitian exchange
matrices by
\begin{eqnarray}
\label{eq:parametrization}  {J^{(1)}} &=& \hat{J}, \qquad {J^{(2)}}
=\tau^x \hat{J} \tau^x, \qquad
  \hat{J}= J (1+\cos(2 \theta) \tau^z  \nonumber \\
&+&\sin(2 \theta)[\cos(\Phi/2) \tau^x-\sin(\Phi/2)
\tau^y]),\end{eqnarray} where
\begin{eqnarray}
\label{eq:parametrization1}
 \theta=|\arctan(t_2/t_1)|,~~~J \sim
\frac{t_1^2+t_2^2}{U}.
\end{eqnarray}
Parity symmetry for the PS amplitudes implies $V^L_L=V^R_R$,
$V^L_R=V^R_L$ (${\rm{Im}} V_L^R=0$). We can estimate
\begin{equation} \label{eq:V} V^L_L \sim
\frac{t_1^2+t_2^2}{U},~~~V^L_{R} \sim \frac{t_1 t_2}{U}  \cos
\frac{\Phi}{2}+ c' \frac{(t_1^2+t_2^2 )t_{12}}{U^2}.
\end{equation}

It is convenient to define even and odd channels $ \psi_{ e,o }
=\frac{\psi_{ L }\pm \psi_{ R }}{\sqrt{2}}$, in terms of which the
parity transformation reads $\psi_{e} \rightarrow \psi_{e},\qquad
\psi_{o} \rightarrow -\psi_{o}$. The most general form of $H_{K} +
H_{PS}$ consistent with parity is
\begin{eqnarray}
\label{eq:Hparity} H_K &=&  J_e {\psi^\dagger}^{e \alpha}
\frac{\vec{\sigma}_\alpha^\beta}{2} \psi_{e
\beta}\cdot(\vec{S}_1+\vec{S}_2)\nonumber \\
+&J_0&{\psi^\dagger}^{o \alpha} \frac{\vec{\sigma}_\alpha^\beta}{2}
\psi_{o \beta}\cdot(\vec{S}_1+\vec{S}_2)\nonumber
\\
&+&[J_m {\psi^\dagger}^{e \alpha}
\frac{\vec{\sigma}_\alpha^\beta}{2} \psi_{o
\beta}+h.c.]\cdot(\vec{S}_1-\vec{S}_2), \nonumber \\
H_{PS} &=&V_e {\psi^\dagger}^{e \alpha} \psi_{e \alpha}+V_o
{\psi^\dagger}^{o \alpha} \psi_{o \alpha}.
\end{eqnarray}
Indeed using Eq.~(\ref{eq:parametrization}) we obtain $H_{K}+H_{PS}$
in this form with
\begin{eqnarray}
\label{eq:phim}
J_{e,o} &=& \frac{\hat{J}^L_L+\hat{J}^R_R \pm(\hat{J}_L^R+\hat{J}_R^L)}{2} = J[1 \pm \sin(2 \theta) \cos(\Phi/2)],\nonumber \\
J_m&=&
\frac{\hat{J}^L_L-\hat{J}^R_R+\hat{J}_L^R-\hat{J}_R^L}{2}\nonumber
\\
&=& J[\cos(2 \theta)- i \sin(2 \theta) \sin(\Phi/2)] = |J_m| e^{i
\phi_m},\nonumber \\
V_{e,o} &=& \frac{V^L_L+V^R_R \pm(V_L^R+V_R^L)}{2},
\end{eqnarray}
where
\begin{equation}
\label{eq:phim1} \phi_m=-\arctan\bigl( \tan(2 \theta) \sin (\Phi/2)
\bigr).
\end{equation}
For finite flux $J_m$ has an imaginary part. To recover real
coupling constants in $H_{K}$ we remove this phase by a
redefinition of the fields
\begin{eqnarray}
\label{eq:rotation} \psi_{e } \rightarrow  \psi_{e
}'= e^{- i \phi_m/2}\psi_{e }, \nonumber \\
\psi_{o } \rightarrow  \psi_{o }'= e^{i \phi_m/2}\psi_{o }.
\end{eqnarray}
In the $\psi'_{e,o}$ basis, $H_K$  has real coupling constants,
$J_e$, $J_o$ and $|J_m|$, and it corresponds to the notation in
Ref.~[\onlinecite{Affleck95}]. It is convenient to define $\psi'_{ 1
} =\frac{\psi'_{ e }+\psi'_{ o }}{\sqrt{2}}$, $ \psi'_{ 2 }
=\frac{\psi'_{e }-\psi'_{ o }}{\sqrt{2}} $. Equivalently, the fields
$\psi'_j$ $(j=1,2)$ are related to the L-R basis by the rotation
$\psi_{i } = (\mathcal{M} e^{ i \tau^z \phi_m/2} \mathcal{M})_i^j
\psi'_{j }$, where $\mathcal{M}=\frac{\tau^z+\tau^x}{\sqrt{2}}$.

As will be discussed in Sec.(\ref{se:observability}), observability
of the QCP in this system is restricted to the regime $t_1 \gg t_2$,
or equivalently small $\theta$ (see
Eq.~(\ref{eq:parametrization1})). In this limit the two impurity
Kondo physics is especially transparent, since each QD is coupled
essentially to one lead.\cite{Sela08} $K$ can be tuned by means of
$t_{12}$. For $K \gg K_c$ the impurities are locked into a singlet,
while for $K=0$ each impurity is Kondo-screened by the nearby lead.
In the case of exact P-H symmetry, occurring for $V_{i}^j=0$, those
points in the $K$-parameters space are separated by a QCP at a
critical value $K=K_c \sim T_K$.\cite{Jones88}

\section{Conductance at the fixed point}
In Ref.~\onlinecite{Sela08} the conductance of a series double QD
was calculated using the tunneling current operator. In this paper
until Sec.~(\ref{se:V}) we use the Kubo linear conductance formula
written in terms of {\it bulk} current correlation function. The
reason for taking this different approach here is that it relates
the conductance to correlation functions in certain field theories,
that can be addressed using boundary conformal field theory or
integrability methods. This allow us to express the conductance of
double QDs described by the 2IKM in terms of correlation functions
in the boundary Ising field theory.

\begin{figure}[h]
\begin{center}
\includegraphics*[width=90mm]{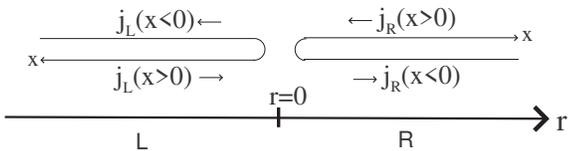}
\caption{Illustration of the physical coordinate $r$ running from
left to right leads and the fictitious coordinate $x$ labeling
position of the chiral fields $\psi_{i \alpha}(x)$ in lead
$i=L,R$.\label{fg:2}}
\end{center}
\end{figure}

\label{se:T=0} The linear conductance can be calculated from the
Kubo formula
\begin{eqnarray}
G=\lim_{L \rightarrow \infty} \lim_{\omega \rightarrow 0}
\frac{e^2}{\hbar \omega(2 L)^2} \int_{- L}^L dr \int_{- L}^L dr'
\int_{- \infty}^\infty d \tau e^{-i \omega \tau} \nonumber \\
\times  \langle J(r,\tau) J(r',0)  \rangle.  \nonumber
\end{eqnarray}
Here $r$ is the physical coordinate; see Fig.~(\ref{fg:2}). It
should be distinguished from the fictitious coordinate $x$ labeling
the chiral fermions $\psi_{i \alpha}(x)$. We define the chiral
current densities in each lead $j_L(x) ={ \psi^\dagger}^{L \alpha}
(x) \psi_{L \alpha}(x)$, $j_R(x) ={ \psi^\dagger}^{R \alpha} (x)
\psi_{R \alpha}(x)$. The bulk current operator $J(r)$ can be written
as
\begin{equation}
J(r) =- \left \{
\begin{array}{ll}
j_R(r)-j_R(-r) & r>0  \\
j_L(r)-j_L(-r) & r<0
\end{array} \right. . \nonumber
\end{equation}
It is useful to define the odd current $j_o(x) = j_L(x)-j_R(x)$,
since $ \int_{-L}^L dr J(r)  = \int_{-L}^L dx j_o(x) {\rm{sgn}}(x
)$. The conductance is given in terms of the odd current correlator,
\begin{eqnarray}
\label{eq:Gjodd} G=\lim_{L \rightarrow \infty} \lim_{\omega
\rightarrow 0} \frac{e^2}{\hbar \omega(2 L)^2} \int_{- L}^L dx
\int_{- L}^L dy \int_{- \infty}^\infty d \tau e^{-i
\omega \tau} \nonumber \\
\times \langle  j_o(x,\tau) j_o(y,0) \rangle {\rm{sgn}}(x y).
\end{eqnarray}

The odd current $j_o(x) ={\psi^\dagger}^{j \alpha} (\tau^z)^j_j
\psi_{j \alpha}$ corresponds to the $z-$component of the flavor
current of the fermions in the L-R basis. We define the flavor
current in terms of the fermions $\psi'_{j \alpha}$ after the
rotation Eq.~(\ref{eq:rotation}),
\begin{equation}
\label{eq:flavor} \vec{j}^{f} =   {{\psi'}^\dagger}^{i \alpha}
\frac{{\vec{\tau}}^j_i}{2} \psi'_{j
\alpha}.~~~(\rm{repeated~indices~summed})
\end{equation}
The transformation Eq.~(\ref{eq:rotation}) amounts to a rotation in
the flavor sector,
\begin{eqnarray}
\label{eq:flavorzx} j_o = 2 [\cos \phi_m({j}^{f})^z -\sin
\phi_m({j}^{f})^y].
\end{eqnarray}

Consider the weak coupling limit $J \rightarrow 0$. It corresponds
to a trivial boundary condition (BC) $ \psi_L(x=0^+) =
\psi_L(x=0^-)$, $ \psi_R(x=0^+) = \psi_R(x=0^-)$, describing free
fermions with full reflection at the boundary. Also this BC makes
apparent the continuity of the chiral fields $\psi_L$, $\psi_R$ at
$x=0$. Accordingly, the odd current correlator is given by
\begin{equation}
\langle j_o(x,\tau) j_o(y,0) \rangle_{J=0} =-\frac{1}{\pi^2} \frac{
1}{(\tau+i (x-y))^2}.  \nonumber
\end{equation}
To calculate the odd current correlator at the nontrivial fixed
point, we apply CFT methods and the Bose-Ising representation used
in Ref.~[\onlinecite{Affleck95}]. In this representation the 4
fermions $\psi_{i \alpha }'$ are represented using a coset
construction in terms of three Wess-Zumino-Witten (WZW) nonlinear
$\sigma$ models, $SU(2)_1^{charge 1}\times SU(2)_1^{charge 2}\times
SU(2)_2^{spin} $, together with a $\mathcal{Z}_2$ Ising model. The
currents of the two $SU(2)_1$ $\sigma$ models are associated with
the charge of each species $\psi'_{1 \alpha}$ and $\psi'_{2
\alpha}$. The current of the $SU(2)_2$ model is associated with the
total spin.

Following Ref.~[\onlinecite{Affleck95}] one may write down
representations of the various operators in the free fermion theory
as product of charge (or isospin) bosons, the total spin boson, and
the Ising field. The $k=2$ WZW model has primary fields of spin
$j=0$ (identity operator $\mathbf{1}$), $j=1/2$ (fundamental field
$g_\alpha$), and $j=1$ (denoted $\vec{\phi}$). The $k=1$ WZW model
only has the identity operator and the $j=1/2$ primary, $h_A$. Their
scaling dimension is given by $\Delta=\frac{j(j+1)}{2+k}$. The Ising
model has three primary fields: the identity operator $\mathbf{1}$
($\Delta=0$), the Ising order parameter $\sigma$ ($\Delta=1/16$),
and the energy operator $\epsilon$ ($\Delta=1/2$). For example the
fermion field is written in this representation as
\begin{equation}
\label{eq:bosonization} \psi'_{i \alpha} \propto (h_i)_1 g_\alpha
\sigma.
\end{equation}
The three factors have dimensions which add correctly to $1/2$. The
representation of other operators can be determined using the
operator product expansion (OPE). For the Ising model the OPE gives
\begin{equation}
\sigma \times \sigma\rightarrow \mathbf{1}+\epsilon, \qquad \sigma
\times \epsilon \rightarrow \sigma, \qquad \epsilon \times \epsilon
\rightarrow \mathbf{1}.  \nonumber
\end{equation}
This OPE is equivalent to that of the $k=2$ WZW model with the
identifications $ \sigma \leftrightarrow g$ and $\epsilon
\leftrightarrow \vec{\phi}$.

Using the OPE, symmetry considerations, and consistency of scaling
dimensions, we shall determine the representation of the odd current
$j_o$. The latter is related in Eq.~(\ref{eq:flavorzx}) to the
flavor current operators $(j^{f})^z$ and
$({j}^{f})^y=\frac{({j}^{f})^+-({j}^{f})^-}{2i }$. First consider
$({j}^{f})^z = \frac{1}{2} ( {{\psi'}^\dagger}^{1\alpha} \psi'_{1
\alpha}- {{\psi'}^\dagger}^{2\alpha}  \psi'_{2 \alpha})$. This is
just the charge difference between flavors, represented by $I^z_1 -
I^z_2$, where $\vec{I}_i$ is the $SU(2)_1^{chargei}$ current,
($i=1,2$). For the operator $({j}^{f})^+ =
{{\psi'}^\dagger}^{1\alpha} \psi'_{2 \alpha}$, we use
Eq.~(\ref{eq:bosonization}),
\begin{eqnarray}
{{\psi'}^\dagger}^{1 \alpha}(x) \psi'_{2 \alpha}(x) \propto \lim_{x'
\rightarrow x} \nonumber \\
 {g^\alpha}^\dagger(x') g_\alpha(x)  (h_1)^{1
\dagger}(x') (h_2)_1(x)  \sigma(x') \sigma(x).  \nonumber
\end{eqnarray}
Consider the OPE of the fundamental field  $g \times  g = \mathbf{1}
+\vec{\phi}$. The operator under consideration is a spin singlet,
ruling out $\vec{\phi}$ in the OPE. To account for the consistency
of scaling dimensions we must have $\sigma \times \sigma \rightarrow
\epsilon$. Hence $ {{\psi'}^\dagger}^{1 \alpha} \psi'_{2 \alpha}
\propto (h_1)^{1 \dagger} (h_2)_1 \epsilon$. The Bose-Ising
representation of the flavor current is summarized in the first
column of Table~\ref{tb:1}. In the second column we consider an
alternative $SO(8)$ representation introduced in Sec.~\ref{se:V}.

\begin{table}
\begin{tabular}{c@{\extracolsep{1.5em}}cc}
\hline \hline
$~$ & $SU(2)_1\times SU(2)_1\times SU(2)_2\times \mathcal{Z}_2$ & $SO(8)$ \\
\hline \hline $({j}^{f})^z$
    & $I^z_1-I^z_2$ & ${\psi^\dagger}_{f} {\psi}_{f}$   \\
\hline
 $({j}^{f})^+$
    & ${(h_1)^1}^\dagger (h_2)_1 \epsilon$ &${\psi^\dagger}_f \chi_2^X$         \\
\hline \hline
\end{tabular}
\caption{\label{tb:1}%
Bose-Ising- versus $SO(8)$ Majorana- representation of the flavor
current.}
\end{table}
The main result of Ref.~[\onlinecite{Affleck95}] is that the
nontrivial BC of the 2IKM at $K=K_c$ corresponds to a change in the
boundary condition occurring only in the Ising sector of the theory:
the nontrivial BC of the electrons corresponds to the free BC on the
Ising chain, whereas the trivial BC for the electrons corresponds to
the Ising model with a fixed boundary spin.~\cite{Cardy89} We shall
refer sometimes to the BC of the full system at the nontrivial fixed
point by ``free" and at the trivial free fermion fixed point by
``fixed".

The remaining sectors of the theory other than the Ising model
remain unaffected. Correlation functions of factors belonging to
sectors other than the Ising model have the form dictated by
conformal invariance, $\langle \mathcal{O}_\Delta(x,\tau)
\mathcal{O}_\Delta(y,0)\rangle = \frac{1}{(\tau+i (x-y))^{2
\Delta}}$, where $\Delta$ is the scaling dimension of $\mathcal{O}$.
This form remains valid both at the trivial and nontrivial fixed
points. On the other hand correlation functions of fields from the
Ising sector do depend on BC. There is a general formula for
correlation function of primary operators for a BC obtained by
fusion with a primary $a$,~\cite{Cardy91,Affleck93}
\begin{equation}
\label{eq:Cardy} \langle \mathcal{O}_\Delta(x,\tau)
\mathcal{O}_\Delta(y,0) \rangle =\frac{1}{(\tau+i (x-y))^{2\Delta}}
\times \left \{
\begin{array}{ll}
1 & x y >0  \\ \frac{S^\Delta_a/S_0^\Delta}{S_a^0/S_0^0}
 & x y <0
\end{array} \right. .
\end{equation}
Here $S_j^a$ are elements of the modular $S-$matrix. For the Ising
model this is given by
\begin{equation}S=
\left(%
\begin{array}{ccc}
 1/2&1/2&1/\sqrt{2} \\
  1/2&1/2&-1/\sqrt{2} \\
  1/\sqrt{2}&-1/\sqrt{2}&0
\end{array}%
\right),  \nonumber
\end{equation}
where the first, second and third rows and columns are labeled by
the fields with scaling dimension $0,1/2,1/16$, respectively. The
change in BC in the 2IKM from trivial to nontrivial fixed points
corresponds to fusion with the spin operator in the Ising
sector.~\cite{Affleck95} Setting $\Delta=1/2, a=1/16$ we have
$\frac{S^\Delta_a/S_0^\Delta}{S_a^0/S_0^0}=-1$. Hence
Eq.~(\ref{eq:Cardy}) gives
\begin{equation}
\label{eq:epepfree}
 \langle \epsilon(x,\tau)
\epsilon(y,0) \rangle_{free} =\langle \epsilon(x,\tau) \epsilon(y,0)
\rangle_{fixed} \cdot {\rm{sgn}}(x y).
\end{equation}
where up to a normalization factor $\langle \epsilon(x,\tau)
\epsilon(y,0) \rangle_{fixed} \propto \frac{ 1}{\tau+i (x-y)}$. We
may interpret this as a phase shift of $\pi/2$ that the energy
operator $\epsilon(x)$ undergoes at $x=0$. We proceed to evaluate
the odd current correlation function. Since the crossed terms
$\langle ({j}^{f})^z ({j}^{f})^y \rangle$ vanish, we obtain
\begin{eqnarray}
\label{eq:oddcorrelation}
 \langle  j_o(x,\tau)
j_o(y,0) \rangle_{free} =\langle  j_o(x,\tau) j_o(y,0) \rangle_{J=0}
 \nonumber \\
\times \left \{
\begin{array}{ll}
1& x y >0  \\
\cos(2 \phi_m) & x y <0
\end{array} \right. .
\end{eqnarray}
One can use the Kubo formula Eq.~(\ref{eq:Gjodd}) to calculate the
conductance. However a calculation is unnecessary: Curiously, one
obtains exactly the same result for the odd current correlation
function, Eq.~(\ref{eq:oddcorrelation}), assuming free fermions with
partially transmitting BC,
\begin{eqnarray}
\label{eq:thetaBC} \psi_L(0^+) = \cos( \phi_m) \psi_L(0^-)+i  \sin
(\phi_m) \psi_R(0^-),
\nonumber \\
\psi_R(0^+) = i  \sin (\phi_m) \psi_L(0^-)+\cos (\phi_m)
\psi_R(0^-).
\end{eqnarray}
This BC corresponds to transmission probability $\sin^2 \phi_m$ per
spin. From Landauer formula~\cite{Landauer57} the linear conductance
at the nontrivial fixed point is
\begin{equation}
\label{eq:G0} G_0 = \frac{2 e^2}{h} \sin^2 \phi_m,
\end{equation}
where $\phi_m$ is given in Eq.~(\ref{eq:phim1}). This gives $G_0=0$
for $\phi_m=0$, in particular for $\Phi=0$, and for the case of a
series QD (in this section $V_{L}^R=0$).\cite{Sela08} Also in the
experimentally relevant regime $t_1 \gg t_2$ (see
Sec.~\ref{se:observability}), corresponding to small $\theta$ [see
Eq.~(\ref{eq:parametrization1})], $G_0 \ll 2e^2/h$.

[Note however, that the actual BC at the nontrivial fixed point
written in terms of the true fermions is very different from
Eq.~(\ref{eq:thetaBC}). This is apparent from the vanishing of the
one-particle $S-$matrix.~\cite{Affleck95} The auxiliary fermions
satisfying linear boundary condition emerge in the $SO(8)$
representation that we shall use in Sec.~(\ref{se:V}).]

\subsection{$T=0$ phase diagram}
\label{se:MAP} 
\begin{figure}[h]
\begin{center}
\includegraphics*[width=60mm]{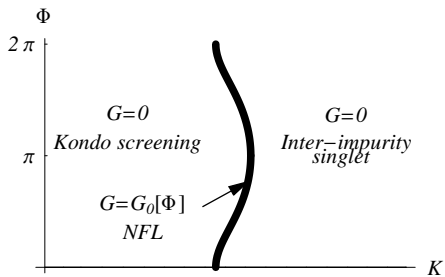}
\caption{Schematic phase diagram as function of inter-impurity interaction $K$
(controlled by $t_{12}$) and flux $\Phi$ for fixed asymmetry (we
assume a generic situation $t_1 \ne t_2$). Here P-H symmetry is
assumed. The conductance is finite only at the NFL curve defined by
$K = K_c[\Phi]$ except for $\Phi=0,2\pi$. \label{fg:2a}}.
\end{center}
\end{figure}

Having found the conductance at the QCP at $K = K_c$,
we shall consider the surrounding FL fixed points and draw a phase
diagram. Here and until Sec.~(\ref{se:HQCP}) we consider the P-H
symmetric model. In this model charge transfer between the leads
leading to finite current is allowed by the exchange interaction in
$H_{K}$. The main role of $\theta = \arctan(t_2/t_1)$ and flux
$\Phi$ is to modify the crossover scales $T_K$ and $K_c$. We plot in
Fig.~(\ref{fg:2a}) the phase diagram at fixed $\theta$ as function
of $K$ and flux. The NFL state occurs along the curve $K =
K_c[\Phi]$, where $K_c \sim T_K[J_e,J_o,J_m]$ and $J_e,J_o,J_m$
depend on flux through Eqs.~(\ref{eq:phim}). This curve is
characterized by a finite conductance $G=G_0$ at $\Phi \ne 0,2 \pi$.
It separates the $K>K_c$ ''local singlet'' phase from the $K<K_c$
Kondo-screened phase.

The conductance vanishes in both FL phases. At $K>K_c$ the system
remains in its weak coupling limit, corresponding to weakly
transmitting tunnel junctions. At $K<K_c$ a Kondo-screened phase is
developed and the two channels $\psi_e$ and $\psi_o$ participate in
the screening of the combined spin$-1$ impurity. In the effective FL
description both the even and odd channels acquire a phase shift of
$\delta_e=-\delta_o=\pi/2$. The conductance vanishes as a result of
\emph{destructive interference} between even and odd channels: an
incoming electron from the left lead $\psi_L^{in} =
\frac{\psi_e^{in}+\psi_o^{in}}{\sqrt{2}}$ scatters into the outgoing
state $ \frac{\psi_e^{out}e^{2 i \delta_e}+\psi_o^{out}e^{2 i
\delta_o}}{\sqrt{2}} =- \psi_L^{out} $ in the left lead,
corresponding to full reflection. (The situation is reversed if a
$\pi/2$ phase shift occurs only in one channel). We point out that
when we include P-H symmetry breaking the conductance is finite in
the FL phases.

In the Hamiltonian Eq.~(\ref{eq:Hparity}) the condition $J_m \ne 0$
is required to mix the impurity singlet and triplet subspaces of the
Hilbert space. At $J_m=0$ the transition at $K=K_c$ corresponds to a
level crossing between those subspaces. We point out that this
special situation occurs in our system for the symmetric point $t_1
= t_2$, ($\theta =\pi/4$), and at zero flux. In this case, when
$K<K_c$ the conductance is $G= \frac{2 e^2}{h}$, since the odd
channel is decoupled, $\delta_o=0$, and as a result of Kondo effect
in the even channel $\delta_e=\pi/2$.

\section{Universal crossover as function of inter-impurity interaction $K$}
\label{se:Tne0} 
\begin{figure}[h]
\begin{center}
\includegraphics*[width=90mm]{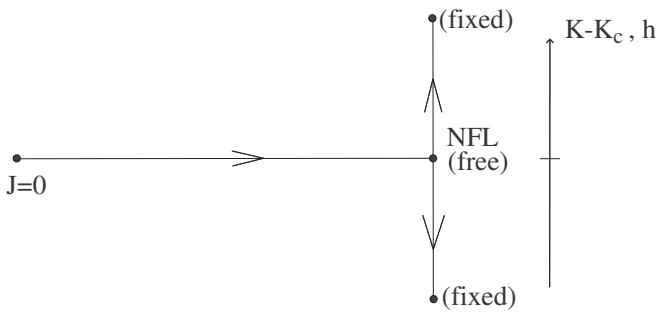}
\caption{Schematic flow diagram of the P-H symmetric 2-impurity
Kondo model. \label{fg:3}}
\end{center}
\end{figure}

In the previous section we calculated the
conductance at the critical value of the inter-impurity exchange
interaction $K=K_c$ and assuming P-H symmetry. In this situation the
system flows from weak coupling $(J=0)$ to a NFL fixed point,
corresponding to free BC in the Ising sector. At finite $|K-K_c|$
the system flows to another fixed points as illustrated in
Fig.~(\ref{fg:3}). Depending on the sign of $K-K_c$, those two
states correspond to fixing the boundary spin in a semi-infinite
Ising chain to point up or down. Note that whereas both the
attractive and $J=0$ fixed points in Fig.~(\ref{fg:3}) correspond to
fixed boundary condition in the Ising sector, they differ by the
impurity spin states. The latter are decoupled and contribute to the
ground state degeneracy only at the repulsive $J=0$ fixed point.

The crossover along the horizontal line in Fig.~(\ref{fg:3}) is
governed by the Kondo energy scale $T_K$. The universality of this
crossover between fixed and free boundary conditions in the Ising
sector, however, is spoiled by irrelevant operators emerging from
other sectors of the theory.

The crossover along the vertical line in Fig.~(\ref{fg:3}) is
governed by an energy scale $T^* =c_1 \frac{(K-K_c)^2}{T_K}$, where
$c_1$ is a numerical factor of $O(1)$. It was argued in
Ref.~[\onlinecite{Affleck95}] that this crossover is in the
universality class of the crossover from free- to fixed-boundary
conditions in the Ising model, driven by a local magnetic field $h$
at the boundary of a quantum Ising chain. The magnetic field $h$ is
linearly related to $K-K_c$,
\begin{equation}
\label{eq:Tstar0} T^*=c_1 \frac{(K-K_c)^2}{T_K} =h^2. \end{equation}
When $T^* \ll T_K$ we may safely ignore sectors other than the Ising
sector in the low energy crossover. This mapping of the 2IKM to the
boundary Ising model opens the possibility to calculate the full
crossover formula for the conductance as function of $h \sim
\frac{K-K_c}{\sqrt{T_K}}$ at finite temperature due to exact
solvability of the boundary Ising model.

\subsection{Boundary Ising model}
It is well known that the scaling limit of the two dimensional
classical Ising model at its bulk critical point is described by a
free massless Majorana field theory. Here we consider the two
dimensional model with a boundary, which is equivalent to the
quantum semi-infinite chain. After unfolding the model in the
standard fashion~\cite{Affleck91} we obtain a left moving Majorana
fermion on the infinite line,
\begin{equation}
\label{eq:HIsing} H_{Ising} = \frac{1}{2}\int_{- \infty}^\infty dx
\chi(x) i
\partial_x \chi(x)+H_B , \qquad H_B= h \sigma_B.
\end{equation}
At $h=0$ this model corresponds to free BC, expressed by the
continuity of the chiral Majorana fermion field $\chi(x)$ at $x=0$.
$h$ is an external magnetic field acting on the boundary spin
$\sigma_B$ only. Clearly $h = \pm \infty$ implies fixed BC. The
boundary spin can be written as~\cite{Ghoshal94}
\begin{equation}
\label{eq:sigma} \sigma_B = i \chi(x=0) a.
\end{equation}
Here $a$ is an additional Majorana fermionic boundary degree of
freedom which anticommutes with $\chi$ and satisfies $a^2=1/2$.

The bulk energy operator of the Ising model corresponds to a mass
term $m \chi \bar{\chi}$, which is a product of a left- and a
right-moving Majorana fields. Therefore the left moving factor of
the energy operator, which is the field we refer to as the energy
operator, is just the free Majorana fermion $\epsilon(x) \sim
\chi(x)$ with dimension $\Delta= 1/2$. Note that $\chi$ was
introduced most naturally within free BC, while $\epsilon$ was
introduced to represent free fermions at the fixed BC fixed point.
Indeed, for free BC of the Ising model $\chi(x)$ is continuous and
$\epsilon(x)$ undergoes a $\pi/2$ phase shift at $x=0$ [see
Eq.~(\ref{eq:epepfree})]. Hence,
\begin{equation}
\label{eq:minus1}
 \epsilon(x) ={ \rm{sgn}}(x) \chi(x).
\end{equation}

\subsection{Energy correlator at finite boundary field}
In a bulk CFT a typical local operator is a product of left and
right moving factors $\phi(x) = \phi_L(x) \phi_R(x)$ where we
suppress the time variable. The Ising model has three primary bulk
operators denoted $\mathcal{O}_{\Delta}$, $(\Delta=1/2,1/16,0)$. In
the presence of a boundary at $x=0$ one can formulate the theory in
terms of left moving fields only, $\phi(x) = \phi_L(x) \phi_L(-x)$,
$x>0$. For example, $\mathcal{O}_{1/2}(x) = \epsilon(x)
\epsilon(-x)$. In particular, at $\tau=0$, $y=-x$ the correlator of
the left moving Ising fields $\epsilon$ at any $h$ is related to the
\emph{one-point function} of the bulk energy operator of the
boundary Ising model, $\langle \epsilon(x,0) \epsilon(-x) \rangle_h
= \langle \mathcal{O}_{1/2}(x) \rangle_h$. The one-point function of
the bulk energy operator was calculated using the integrability of
the boundary Ising model with the
result\cite{Ghoshal94,Konic96,LeClair96}
\begin{equation}
\label{eq:konikform} \langle \mathcal{O}_{1/2}(x)
\rangle_h=\int_{-\infty}^\infty \frac{ du}{2 \pi}\frac{e^{2 iu
x}}{1+e^{\beta u}}\frac{i h^2 /2-u}{i  h^2 /2+u}.
\end{equation}
Here $\beta =T ^{-1}$ is the inverse temperature. More generally
consider the correlation function $\mathcal{C}_h(x,y,\tau) = \langle
\epsilon(x,\tau) \epsilon(y,0) \rangle_h$. Consider a perturbative
calculation of $\mathcal{C}_h(x,y,\tau)$ in $H_B$. It can be
shown~\cite{Affleck93} that (i) the correction vanishes for $xy>0$,
(ii) for $x y<0 $ the correction is a function of $z = \tau+i(x-y)$.
This implies that we can \emph{analytically continue} the one-point
function to find $\mathcal{C}_h(x,y,\tau)$,
\begin{equation}
\label{eq:1pf} \mathcal{C}_h(x,y,\tau)= \langle \mathcal{O}_{1/2}(x)
\rangle_h |_{x \rightarrow - iz}, \qquad x>0,y<0. \end{equation} For
$x<0,y>0$ one can use $\mathcal{C}_h(x,y,\tau) =
-\mathcal{C}_h(y,x,-\tau)$, where the $-$ sign arises from the
fermionic nature of $\epsilon$.

\subsection{Direct calculation of the energy correlator}
For the present problem the desired correlator can be computed
directly as will be done in this subsection. We turn to a
calculation of the Majorana Green function (GF)
$\mathcal{G}(\tau,x,y) =- \langle \chi(x,\tau) \chi(y,0) \rangle $
at finite $h$ and temperature $T = \beta^{-1}$. From
Eq.~(\ref{eq:minus1}), the energy correlator is
\begin{equation}
\label{eq:minus} \langle \epsilon(x,\tau) \epsilon(y,0) \rangle_h
=-\mathcal{G}(\tau,x,y){ \rm{sgn}}(x y).
\end{equation}
For $h=0$,  $\mathcal{G}(\tau,x,y)$ is a free fermion GF,
\begin{eqnarray}\mathcal{G}^{(0)}(\tau,x,y) =\frac{1}{2 \pi} \frac{-\pi/\beta}{\sin
(\frac{\pi}{\beta}(\tau+i(x-y)))} \nonumber \\= \frac{1}{\beta}\sum_n e^{-i \omega_n \tau} \mathcal{G}^{(0)}(i \omega_n,x,y) \nonumber \\
 = \frac{i
}{\beta}\sum_n e^{-i \omega_n (\tau+i(x-y))} \nonumber \\
\times[\theta(- \omega_n)\theta(x-y)-\theta( \omega_n)\theta(y-x)],
\nonumber
\end{eqnarray}
where $\omega_n=\frac{\pi}{\beta}(1+2 n)$. Since the interaction in
Eq.~(\ref{eq:sigma}) is quadratic in fermion fields, we may sum up
the perturbation series in the boundary magnetic field exactly,
giving
\begin{eqnarray}
\label{eq:epsilonepsilon} \mathcal{G}(i \omega_n,x,y) =
\mathcal{G}^{(0)}(i \omega_n,x,y) \nonumber \\
+h^2
\mathcal{G}^{(0)}(i \omega_n,x,0) \mathcal{G}_a(i \omega_n)
\mathcal{G}^{(0)}(i \omega_n,0,y).
\end{eqnarray}
Here $ \mathcal{G}_a(i \omega_n) =-\int_0^\beta d \tau e^{i \omega_n
\tau} \langle a(\tau) a \rangle$ is the $a$ propagator. When $a$ is
decoupled, its propagator is given by $ \mathcal{G}_a^{(0)}(i
\omega_n) = (i \omega_n)^{-1}$. Eq.~(\ref{eq:epsilonepsilon})
becomes exact when $ \mathcal{G}_a(i \omega_n) $ is calculated to
infinite order in $h$. This is accomplished by the self energy
$\Sigma_a(i \omega_n) = h^2 \mathcal{G}^{(0)}(i \omega_n,0,0) =  -i
 h^2 {\rm{sgn}}(\omega_n)/2$. Thus
\begin{equation}
\label{eq:Ga} \mathcal{G}_a(i \omega_n) = \bigl( i \omega_n + i h^2
{\rm{sgn}}(\omega_n)/2 \bigr)^{-1}. \end{equation} Plugging this
result in Eq.~(\ref{eq:epsilonepsilon}) yields the result
\begin{eqnarray}
\label{eq:epsilonMatzubara} &&\mathcal{G}(i \omega_n,x,y) =
\mathcal{G}^{(0)}(i \omega_n,x,y)+i
 e^{\omega_n(x-y)}  \nonumber\\
&\times&\sum_{s=\pm 1} \theta(s \omega_n)    \theta(s y) \theta(-s
x)\frac{ h^2}{ \omega_n+ h^2 {\rm{sgn}}( \omega_n)/2}.
\end{eqnarray}
When $ x y >0$ there is no dependence on $h$. To compare
Eq.~(\ref{eq:epsilonMatzubara}) in the nontrivial region $xy <0$
with the result obtained by analytic continuation of the one point
function of the energy operator, Eq.~(\ref{eq:konikform}), we write
the Fourier transform of Eq.~(\ref{eq:epsilonMatzubara}) into
\begin{equation}
\mathcal{G}(\tau,x,y)=\int_{-\infty}^\infty \frac{du}{2 \pi}
\frac{e^{u(\tau+i(x-y))}}{1+e^{\beta u}}\frac{i h^2
{\rm{sgn}}(x-y)/2-u}{i  h^2 {\rm{sgn}}(x-y)/2+u}, \nonumber
\end{equation}
valid for $x \cdot y<0$. One arrives at the same result using
Eqs.~(\ref{eq:konikform}), (\ref{eq:1pf}), and (\ref{eq:minus}). In
this notation the integration variable $u$ is related to the
momentum of the particles used in the form factors method.

\subsection{Finite temperature conductance} At finite temperature the
conductance is obtained by analytic continuation
\begin{eqnarray}
\label{eq:GjoddfiniteT} G=\lim_{L \rightarrow \infty} \lim_{\omega
\rightarrow 0} \frac{i e^2}{\hbar \omega(2 L)^2}
 \int_{- L}^L dx \int_{- L}^L dy ~{\rm{sgn}}(x y) \nonumber \\ \times \int_{-
\beta/2}^{\beta/2} d \tau e^{-i \nu_n \tau} \langle j_o(x,\tau)
j_o(y,0) \rangle \big|_{i \nu_n \rightarrow \omega+i 0^+},
\end{eqnarray}
where $\nu_n=\frac{2 \pi n}{\beta}$. For $h=0$ the finite
temperature odd current correlator, $ \langle j_o(x,\tau) j_o(y,0)
\rangle_{free}$ is given by Eq.~(\ref{eq:oddcorrelation}) where $
\langle j_o(x,\tau) j_o(y,0) \rangle_{J=0} = -\frac{1}{\beta^2
\sin^2 [\frac{1}{\beta}(\tau+i(x-y))]}$. At finite $h$ we use
Eq.~(\ref{eq:flavorzx}) and $\langle (j^f)^z(x,\tau) (j^f)^y(y)
\rangle=0$, leading to
\begin{eqnarray}
\langle  j_o(x,\tau) j_o(y,0) \rangle_h =4 \cos^2 \phi_m \langle
(j^f)^z(x,\tau) (j^f)^z(y) \rangle \nonumber \\
+4 \sin^2 \phi_m \langle (j^f)^y(x,\tau) (j^f)^y(y) \rangle.
\nonumber
\end{eqnarray}
Using the Bose Ising representation of the flavor currents, given in
Table~\ref{tb:1}, and Eq.~(\ref{eq:minus}), we obtain
\begin{eqnarray}
\langle  j_o(x,\tau) j_o(y,0) \rangle_h =-4 \cos^2 \phi_m \left(
\mathcal{G}^{(0)}(\tau,x,y) \right)^2 \nonumber \\
 -4 \sin^2 \phi_m
{\rm{sgn}}(x y)
 \mathcal{G}^{(0)}(\tau,x,y)   \mathcal{G}(\tau,x,y).\nonumber
\end{eqnarray}
Compared to free BC, the odd current correlator obtains an
additional term
\begin{eqnarray}
\label{eq:odd}  \langle  j_o(x,\tau) j_o(y,0) \rangle_h =  \langle
j_o(x,\tau) j_o(y,0) \rangle_{free}
 \nonumber \\
 -2 \sin^2
\phi_m \frac{
\mathcal{G}(\tau,x,y)-\mathcal{G}^{(0)}(\tau,x,y)}{\beta \sin
(\frac{\pi}{\beta}(\tau+i (x-y)))} .
\end{eqnarray}
The first term contributes $\frac{2e^2}{h} \sin^2 \phi_m$ to the
conductance. The second term is nonvanishing only for $x y<0$, as
can be seen from Eq.~(\ref{eq:epsilonMatzubara}). Note that
$\mathcal{G}(\tau,x,y)-\mathcal{G}^{(0)}(\tau,x,y) \rightarrow_{h
\rightarrow \infty} - 2 \mathcal{G}^{(0)}(\tau,x,y) \theta(-x y)$,
hence
\begin{equation}
\langle  j_o(x,\tau) j_o(y,0) \rangle_{h \rightarrow \infty} =
\langle j_o(x,\tau) j_o(y,0) \rangle_{fixed},\nonumber
\end{equation}
as expected. At finite $T$ and $h$ the contribution of the second
term to the conductance is given by $ 2 {\rm{Re}} G_1$ where
\begin{eqnarray}
G_1=\lim_{L \rightarrow \infty} \lim_{\omega \rightarrow 0} \frac{-i
e^2}{\hbar \omega(2 L)^2} \int_{- L}^0 dx \int_{0}^L dy \nonumber \\
\times \langle j_o(x) j_o(y) \rangle^{(1)}_{i \nu_n \rightarrow
\omega+i 0^+},\nonumber
\end{eqnarray}
where $\langle j_o(x) j_o(y) \rangle^{(1)} =\langle j_o(x) j_o(y)
\rangle_h-\langle j_o(x) j_o(y) \rangle_{free} $. This correlator
can be expressed as a Matsubara sum
\begin{eqnarray}
\langle j_o(x<0) j_o(y>0)
\rangle^{(1)}_{i \nu_n } = \frac{4\sin^2\phi_m}{\beta^2}
\int_{-\beta/2}^{\beta/2} d \tau e^{i \nu_n \tau}  \nonumber
\\
\times \sum_{m,l} \theta(\omega_m)\theta(\omega_l) e^{- i
(\omega_m+\omega_l)(\tau+i (x-y))} \frac{ i h^2}{i \omega_l + i
h^2/2}  \nonumber \\
=\frac{\sin^2\phi_m}{\pi^2} \frac{(2 \pi )^2}{\beta} e^{\nu_n (x-y)}
2 \sum_{l=0}^{n-1} \frac{\beta h^2}{2 \pi(1+2 l)+\beta
h^2}.\nonumber
\end{eqnarray}
The sum is evaluated as an analytic function of $\nu_n = \frac{2 \pi
n}{\beta}$ in terms of the digamma function (the logarithmic
derivative of the gamma function) $\psi(z) =d \log\Gamma(z)/dz$,
\begin{eqnarray}
 \sum_{l=0}^{n-1} \frac{4 \pi}{2 \pi(1+2 l)+\beta
h^2}=  \psi(\frac{1}{2}+\frac{\beta h^2}{4 \pi}+\frac{\beta \nu_n}{2
\pi} ) -\psi(\frac{1}{2}+\frac{\beta h^2}{4 \pi}).\nonumber
\end{eqnarray}
Performing the analytic continuation $i \nu_n \rightarrow \omega+i
0^+$, sending $\omega \rightarrow 0$, and performing the spatial
integrations we obtain
\begin{equation}
\label{eq:con} G / G_0 =1- F[T/T^*],\qquad F[t]=\frac{1}{4 \pi t}
{\rm{Re}}~ \psi_1 \left(\frac{1}{2}+\frac{1}{4 \pi t}\right),
\end{equation}
where $\psi_1(z) = d^2 \log\Gamma(z)/dz^2$ is the trigamma function
and $G_0$ and $T^*$ are given in Eqs.~(\ref{eq:G0}) and
(\ref{eq:Tstar0}). The scaling function $F[x]$ has the properties
$F[0]=1$ and $F[\infty] =0$. A signature of a NFL is the existence
of relevant operators in the Hamiltonian with scaling dimension
$\Delta< 1$. The QD setup discussed here allows to observe that the
inter-impurity interaction is such a relevant perturbation with
$\Delta=1/2$. According to Eq.~(\ref{eq:con}) the crossover from $G
\sim G_0$ to insulating FL state as function of $K-K_c$ occurs at a
value of $|K-K_c|$ which scales with temperature as $T^{1/2}$.


\subsection{Conductance in the model
of Zar\'{a}nd \emph{et. al.}$^{17}$}

 \label{eq:Zarand} We
pause here to comment on an application of the Ising model with
boundary magnetic field for a different double QD model proposed by
Zar\'{a}nd \emph{et. al.}~\cite{Zarand06} as a realization of the
2IKM. We will show that in this system the full crossover of the
conductance as function of $K$ in the P-H symmetric point can be
expressed in terms of the one point function of the spin operator of
the boundary Ising model.

\begin{figure}[h]
\begin{center}
\includegraphics*[width=50mm]{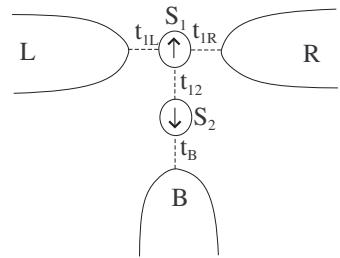}
\caption{Schematic description of the double QD proposed by
Zar\'{a}nd \emph{et. al.}~\cite{Zarand06} as a realization of the
2IKM. \label{fg:zarand}}
\end{center}
\end{figure}

Consider a modified QD system with an additional lead B coupled only
to $S_2$ as in Fig.~(\ref{fg:zarand}). Transport takes place between
the left and right leads, where lead B acts as a screening channel
for spin $S_2$. The analysis of Sec.~\ref{se:T=0} goes through, and
the conductance is given by Eq.~(\ref{eq:Gjodd}), where the odd
current is still written as $j_o(x) ={\psi^\dagger}^{j \alpha}
(\tau^z)^j_j \psi_{j \alpha}$, $j=1,2=L,R$. The next step in
Sec.~\ref{se:T=0} was to rewrite $j_o(x)$ in the basis $\psi'$ which
is natural in the representation of the 2IKM Hamiltonian. For the
present system this basis is
\begin{eqnarray}
\psi_1' = \frac{t_{1L} \psi_L+t_{1R}
\psi_R}{\sqrt{t_{1L}^2+t_{1R}^2}},~~~ \psi_2' = \psi_B.  \nonumber
\end{eqnarray}
A third fermion $ \psi_3' = \frac{-t_{1R} \psi_L+t_{1L}
\psi_R}{\sqrt{t_{1L}^2+t_{1R}^2}}$ is decoupled from the impurities.
We specialize to $t_{1L}=t_{1R}$. In this basis the correlator
$\langle j_o(x,\tau) j_o(y,0) \rangle$ occurring in the Kubo formula
factorizes into the product of GFs for $\psi_1'$ and $\psi_3'$,
where the latter is a free fermion GF. For the GF of $\psi_1'$ we
use the bosonization formula Eq.~(\ref{eq:bosonization}), where the
only factor which is sensitive to the critical point is the Ising
spin operator, leading to
\begin{equation}
\label{eq:jj} \langle j_o(x,\tau) j_o(y,0) \rangle =-
\frac{1}{\pi^2} \frac{\langle \sigma(x,\tau) \sigma(y,0)
\rangle}{(\tau+i(x-y))^{2-1/8}}.
\end{equation}
For finite $h$ and $T$ Eqs.~(\ref{eq:GjoddfiniteT}) and
(\ref{eq:jj}) express the conductance in terms of the 2-point
function for the chiral spin operator at finite magnetic field $h$.
Following the analysis leading to Eq.~(\ref{eq:1pf}), the 2-point
function for the chiral spin operator is related to the one point
function of the bulk spin operator by analytic continuation,
\begin{equation}
\langle \sigma(x,\tau) \sigma(y,0) \rangle_h=
\langle\mathcal{O}_{1/16}(x) \rangle_h |_{x \rightarrow - iz},
\qquad x>0,y<0.  \nonumber
\end{equation}
The calculation of the one point function of the Ising spin at
finite magnetic field $h$ was addressed using integrability and the
form factors method.~\cite{Ghoshal94,Konic96,LeClair96} Different
than the case of the energy operator, a closed expression for
$\langle\mathcal{O}_{1/16}(x) \rangle_h$ is not available. In the
limiting cases $h=0$ and  $h=\pm \infty$ CFT methods can be
used.~\cite{Zarand06} For BC obtained by fusion with operator $a$,
Eq.~(\ref{eq:Cardy}) gives (with $\Delta=1/16$)
\begin{equation}
\label{eq:spin} \langle \sigma(x,\tau) \sigma(y,0) \rangle = \left
\{
\begin{array}{ll}
\frac{ 1}{(\tau+i (x-y))^{1/8}} & x y >0  \\
\frac{ S_{(a)}}{(\tau+i (x-y))^{1/8}} & x y <0
\end{array} \right. ,
\end{equation}
where $S_{(a)}=\frac{S_a^{1/16}/S_0^{1/16}}{S_a^0/S_0^0}$. It is
easy to calculate the conductance with Eq.~(\ref{eq:spin}), with the
result $G=\frac{e^2}{h}(1-S_{(a)})$. At weak coupling $J=0$ ($a=0$)
we have $S_{(0)}=1$, $G=0$. This is also the result for the BC
obtained by starting at the QCP and setting $K>K_c$ (local singlet
phase). At the QCP ($a=1/16$) we have $S_{(1/16)}=0$, $G=e^2/h$. In
the Kondo screened phase ($a=1/2$) we have $S_{(1/2)}=-1$, $G=2
e^2/h$. We leave for a future work to apply Eq.~(\ref{eq:jj}) in
order to interpolate between those values of $G$ at finite
temperature and $h$ $\propto (K-K_c$). The additional difficulty for
this system arises due to the presence of the $\sigma$ GF rather
than the $\epsilon$ GF.

\section{Universal Crossover at finite Potential scattering}
\label{se:HQCP} Until here we assumed P-H symmetry and emphasized
that the crossover is in the universality class of the boundary
Ising model. Now we shall consider the more general situation with
potential scattering (PS). We will see that the Ising and charge
$SU(2)_1$ sectors of the theory are coupled. However this coupling
can be written in a simple quadratic form in the Majorana $SO(8)$
representation that will be introduced below.

It is convenient to write $H_{PS}$, defined in Eq.~(\ref{eq:H}), in
the $\psi'$ basis [defined in Eq.~(\ref{eq:rotation})],
\begin{eqnarray}
\label{eq:HPS} H_{PS}& =& \frac{V_{L}^L+V_{R}^R}{2}
({\psi'}^\dagger_1 {\psi'}_1 + {\psi'}^\dagger_2 {\psi'}_2)
\\&+& {\rm{Re}}V_{L}^R({\psi'}^\dagger_1 {\psi'}_2 +
{\psi'}^\dagger_2 {\psi'}_1)
\nonumber \\
&+&V_A ({\psi'}^\dagger_1 {\psi'}_1 - {\psi'}^\dagger_2 {\psi'}_2)+
V_B i( {\psi'}^\dagger_1 {\psi'}_2 - {\psi'}^\dagger_2 {\psi'}_1),
\nonumber
\end{eqnarray}
where
\begin{eqnarray}
\label{eq:VAB} V_A=\frac{V_L^L-V_R^R}{2} \cos \phi_m
-{\rm{Im}}V_{L}^R \sin \phi_m,
\nonumber \\
V_B=\frac{V_L^L-V_R^R}{2} \sin \phi_m +{\rm{Im}}V_{L}^R \cos \phi_m.
\end{eqnarray}
In the parity symmetric case $V_A=V_B=0$.

At the QCP the PS terms describing charge transfer between channels
${\psi'}_1$ and ${\psi'}_2$ generate relevant
perturbations.\cite{Affleck95} To see this consider their Bose-Ising
representation (using Table~\ref{tb:1})
\begin{eqnarray}
\label{eq:VLR} {\psi'}^\dagger_1 {\psi'}_2 + h.c. \sim (h_1)^\dagger
\tau^z (h_2)
\epsilon, \nonumber \\
i{\psi'}^\dagger_1 {\psi'}_2 + h.c. \sim (h_1)^\dagger  (h_2)
\epsilon.
\end{eqnarray}
At the non-trivial fixed point the energy operator $\epsilon$
``disappears" by double fusion; hence one obtains two relevant
boundary operators $(h_1)^\dagger \tau^z (h_2)$ and $(h_1)^\dagger
(h_2)$, with dimension $\Delta=1/2$. In the parity symmetric case
only the first operator is allowed. These relevant operators have
the dimension of a free fermion. Following Gan\cite{Gan95} a fermion
representation emerges naturally in the SO(8) representation that we
shall introduce in the next subsection. In order for these relevant
operators to have bosonic statistics, in the SO(8) representation
indeed they are written as a product of a bulk fermion with a local
fermion with dimension $\Delta=0$, which can be associated with a
leftover impurity degree of freedom.

On the other hand the intra-channel PS terms lead to marginal
operators at the QCP,
\begin{eqnarray}
\label{eq:VLL} {\psi'}^\dagger_1 {\psi'}_1 \pm {\psi'}^\dagger_2
{\psi'}_2  \sim I^z_1  \pm  I^z_2.
\end{eqnarray}

\subsection{Fixed point Hamiltonian in SO(8) representation}
Following Ref.~[\onlinecite{Maldacena97}] we bosonize the original
theory and introduce four left moving bosonic fields
$:{\psi'^\dagger}^{j \alpha}\psi'_{j \alpha}: = \frac{1}{2 \pi}
\partial_x \phi_{j \alpha}$. In terms of the
bosons we can write the fermions as $\psi'_{j \alpha} \sim F_{j
\alpha} e^{-i \phi_{j \alpha}}$. The Klein factors $F_{i \alpha}$
take care of our sign convention required for products of
exponentials of bosonic fields. They satisfy~\cite{Zarand98}
\begin{eqnarray}
\label{eq:Klein} [F_{\mu}, N_{\nu}]=\delta_{\mu \nu} F_{\mu}, \qquad
\{F_{\mu},F^\dagger_{\nu}\} = 2 \delta_{\mu \nu}, \nonumber \\
(F_{\mu} F^\dagger_{\mu}=F^\dagger_{\mu} F_{\mu}=1),\qquad
\{F_{\mu},F_{\nu}\} =0,
\end{eqnarray}
and $[F_{\mu},\phi_{\nu}]=0$, where $\mu,\nu=\{i,\alpha \}$ and
$N_{\mu}$ is the fermion number of species $\mu$.

 Subsequently 4 linear bosonic combinations are
defined, corresponding to charge, spin, flavor, and difference of
spin between the flavors,
\begin{eqnarray}
\phi_c &=&\frac{1}{2} \sum_{j \alpha} \phi_{j \alpha}, \qquad
~~~\phi_s =\frac{1}{2} \sum_{j \alpha} (\sigma^z)_\alpha^\alpha
\phi_{j \alpha}, \nonumber \\
\phi_f &=&\frac{1}{2} \sum_{j \alpha} (\tau^z)_j^j \phi_{j \alpha},
\qquad \phi_X =\frac{1}{2} \sum_{j \alpha} (\tau^z)_j^j
(\sigma^z)_\alpha^\alpha\phi_{j \alpha}. \nonumber
\end{eqnarray}
Since the exponents of these new bosons have dimension $1/2$, we
define new fermions $\psi_A \sim F_A  e^{-i \phi^A}$, $A=c,s,f,X$.
The new Klein factors satisfy Eq.~(\ref{eq:Klein}) with
$\mu,\nu=c,s,f,X$. To fix a convention we define~\cite{Zarand98}
\begin{eqnarray}
F^\dagger_X F^\dagger_s = F_{1 \uparrow}^\dagger F_{1
\downarrow},~~F_X F_s^\dagger = F^\dagger_{2 \uparrow} F_{2
\downarrow},~~F^\dagger_X F^\dagger_f = F_{1 \uparrow}^\dagger F_{2
\uparrow}.\nonumber
\end{eqnarray}
The free part of the Hamiltonian can be written equivalently in
bosonic or fermionic form,
\begin{eqnarray}
H_0 =\sum_{A} \int \frac{dx}{2 \pi} (\partial_x \phi_A)^2= \sum_{A}
\int dx {{\psi}^\dagger}_A (i
\partial_x) {\psi}_{A}.  \nonumber
\end{eqnarray}
 Taking the
real and imaginary parts of those fermions we obtain 8 Majorana
fermions
\begin{eqnarray}
\chi_1^A=\frac{{\psi^\dagger}_A+\psi_A}{\sqrt{2}},\qquad
\chi_2^A=\frac{{\psi^\dagger}_A-\psi_A}{\sqrt{2}i}. \nonumber
\end{eqnarray}

One can establish a connection between the description of the 2IKM
in terms of $SU(2)_1^{charge 1}\times SU(2)_1^{charge 2}\times
SU(2)_2^{spin} \times \mathcal{Z}_2$ with 8 Majorana fermions. The
two $SU(2)_1$ groups can be represented in terms of two bosons
$\frac{\phi_c \pm \phi_f}{\sqrt{2}}$. The $SU(2)_2^{spin}$ current
$\vec{j}^{s} = \frac{1}{2} {{\psi'}^\dagger}^{i \alpha}
{\vec{\tau}}^\beta_\alpha \psi'_{i \beta}$ has the representation $
(j^{s})^z = {\psi^\dagger}_s\psi_s, (j^{s})^+ =\sqrt{2} \chi_1^X
{\psi^\dagger}_s$. Of particular interest for the present work, the
flavor current Eq.~(\ref{eq:flavor}) has the representation (see
Table \ref{tb:1})
\begin{eqnarray}
\label{eq:flavormajorana} (j^{f})^z = {\psi^\dagger}_{f}\psi_{f},
\qquad (j^{f})^+ =-\sqrt{2} i {\psi^\dagger}_{f} \chi_2^X.
\end{eqnarray}

The Ising fermion $\chi$ can be identified with $\chi_2^X$. In fact
the nontrivial BC involves only one out of the 8 Majorana fermions,
reading $\chi_2^X(0^-) = -\chi_2^X(0^+)$. For a description of the
physics relative to the nontrivial fixed point it is convenient to
work with the continuous Ising fermion field
\begin{equation}
\label{eq:gaugeout} \chi(x) ={\rm{sgn}}(x) \chi_2^X(x)=\epsilon(x)
{\rm{sgn}}(x) .
\end{equation}
Using Eq.~(\ref{eq:sigma}), in the P-H symmetric case the relevant
operator can be written as \begin{equation}
\label{eq:sigmaB}\sigma_B = i \chi(x=0) a = i ({\rm{sgn}}(x)
\chi_2^X(x))_{x=0} \cdot a .
\end{equation}

Now consider the non P-H symmetric case. From the SO(8)
representation of the flavor current Eq.~(\ref{eq:flavormajorana}),
the two PS terms in Eq.~(\ref{eq:VLR}), $( {j^f})^{x,y}$, are
written in the trivial fixed point as $i \chi_2^X \chi_{1,2}^f$. CFT
methods tell us that the operators at the QCP are obtained from the
operators at the trivial fixed point by double fusion with the spin
operator of the Ising model. Having identified $\chi_2^X$ with the
Ising fermion, double fusion gives $\chi_2^X \to 1 +\chi_2^X$. To
obtain the correct bosonic statistics we argue that this fusion rule
should be modified to
\begin{equation}
\chi_2^X \to a +\chi_2^X, \nonumber
\end{equation}
where $a$ is the local fermion appearing in Eq.~(\ref{eq:sigmaB}).
Hence the relevant PS operators at the QCP are
\begin{eqnarray}
\label{eq:chia}
(h_1)^\dagger \tau^z (h_2) \sim i \chi_1^{f} a, \nonumber \\
(h_1)^\dagger (h_2) \sim i \chi_2^{f} a.
\end{eqnarray}
Thus, $a$ couples the Ising sector with the charge sectors. The main
argument in favor of this form is obtained by considering the
self-correlation function of the relevant operators, \emph{e.g.},
\begin{equation}
\langle  (h_1^\dagger h_2)(\tau) (h_1^\dagger h_2) \rangle \sim
\mathcal{G}^{(0)}(\tau,0,0)  \mathcal{G}_a(\tau),\nonumber
\end{equation}
at the P-H symmetric point. Fourier transforming Eq.~(\ref{eq:Ga})
for $\mathcal{G}_a(i \omega_n)$ we can deduce the behavior of
$\mathcal{G}_a(\tau)$: in the limit $\tau \ll h^2(\tau \gg h^2)$,
the correlator $\mathcal{G}_a(\tau)$ goes like $\tau^{0}
(\tau^{-1})$. This implies that in these two limits the correlator
$\langle (h_1^\dagger h_2)(\tau) (h_1^\dagger h_2) \rangle$ goes
like $\tau^{-1} (\tau^{-2})$, respectively, as expected from an
operator with scaling dimension $\Delta = \frac{1}{2}(1)$. This
scaling behavior is obtained relying on the fact that $a$ contains
the information about the crossover. It explains why $a$, and not
some other decoupled local operator, should be coupled to
$\chi_1^{f}$ and $\chi_2^{f}$ in Eq.(\ref{eq:chia}). On the
contrary, the presence of an additional decoupled local operator at
the QCP is ruled out as inconsistent with the ground state
degeneracy. Away from the P-H symmetric point, the local operator
$a$ becomes also sensitive to the deviation from the QCP due to
potential scattering and $\mathcal{G}_a$ is modified relative to
Eq.~(\ref{eq:Ga}).

Putting together Eqs.(\ref{eq:sigmaB}) and (\ref{eq:chia}), the
correction to the fixed point Hamiltonian in SO(8) representation is
\begin{equation}
\label{eq:deltaH} \delta H = i \left(\lambda_1 \chi_2^X(x)
{\rm{sgn}}(x) + \lambda_2 \chi_1^{f}(x) +\lambda_3
\chi_2^{f}(x)\right) a \big|_{x=0},
\end{equation}
with
\begin{eqnarray}
\label{eq:lambdas}
\lambda_1 &=& c_1 \frac{K-K_c}{\sqrt{T_K}}, \nonumber \\
(\lambda_2,\lambda_3)&=&c_2 \sqrt{T_K}\nu ({\rm{Re}}V_L^R,V_B),
\end{eqnarray}
where $V^R_L $ and $V_B$ are given in Eqs.~(\ref{eq:V}) and
(\ref{eq:VAB}), and $c_1$ and $c_2$ are constants of $O(1)$. This
estimate of $(\lambda_2,\lambda_3)$ will be justified below; as we
shall see, based on the dimension $\Delta = 1/2$ of the three
relevant operators in Eq.~(\ref{eq:deltaH}) we obtain the crossover
energy scale
\begin{equation} \label{eq:Tstar}T^* =  \lambda_1^2 +
\lambda_2^2 +\lambda_3^2 \equiv \lambda^2.
\end{equation}

To estimate $\lambda_{2}$ and $\lambda_3$ we consider the
renormalization group flow of the inter-channel potential scattering
operators $( {{\psi'}^\dagger}^{1 \alpha} {\psi'}_{2 \alpha} \pm
h.c.)$. In the presence of those operators the flow to the QCP stops
at energy scale $T^{*}_{LR}$. To estimate $T^{*}_{LR}$ we consider
the renormalization of these operators in the perturbative regime at
energy scales $D \gg T_K$ and then in the nonperturbative regime at
energy scales $D \ll T_K$, respectively.
 (A related calculation for the 2 channel Kondo model appears in
[\onlinecite{Pustilnik04}]). We assume that $K = K_c$. At the
initial scale $D_0 \gg T_K$ the dimensionless bare value of these PS
operators are $\Delta_0 = \nu {\rm{Re}}V_{L}^R$ and $\Delta'_0  =
\nu V_B$; see Eq.~(\ref{eq:HPS}). We assume $\Delta_0,\Delta'_0 \ll
1$. Since in the weak coupling regime potential scattering does not
renormalize, we have
\begin{equation}
\Delta(T_K) \sim \Delta_0, \qquad \Delta'(T_K) \sim
\Delta'_0.\nonumber
\end{equation}
These can be viewed as the initial values of the coupling constants
of the relevant perturbations $(h_1)^\dagger \tau^z (h_2)$ and
$(h_1)^\dagger  (h_2)$, respectively. Since these operators have
dimension $1/2$, the dependence of their coupling constants on $D
\ll T_K$ is described by
\begin{equation}
\frac{\Delta(D)}{\Delta(T_K)} \sim \frac{\Delta'(D)}{\Delta'(T_K)}
\sim \left( \frac{T_K}{D}\right)^{1/2}.\nonumber
\end{equation}
The condition $\max \{ \Delta(T^{*}_{LR}) , \Delta'(T_{PS}) \}\sim
1$ gives the estimate
\begin{equation}
\label{eq:estimate} T^{*}_{LR} \sim \max \{T_K \Delta_0^2,T_K
({\Delta'}_0)^2 \} .
\end{equation}
A more precise estimate would take into account higher order terms
in the $\beta-$function for $\Delta,\Delta'$. However, we expect
that this would only change our estimate of $T^*_{LR}$ by
logarithmic factors.

Identifying $T^*_{LR}$ with $\lambda_2^2 + \lambda_3^2$ in
Eq.~(\ref{eq:Tstar}) gives the estimate for $\lambda_2$ and
$\lambda_3$ given in Eq.~(\ref{eq:lambdas}). Under the condition
$\Delta_0, \Delta'_0 \ll 1$ one has a wide energy range $T^{*}_{LR}
\ll D \ll T_K$ for the observation of the QCP. This can occur in a
certain parameters regime, as we discuss in
Sec.~(\ref{se:observability}).

We point out that our estimate for the energy scale $T^*_{LR}$,
Eq.~(\ref{eq:estimate}), which agrees with [\onlinecite{Zarand06}],
is inconsistent with that of Sakai and Shimizu, who studied the 2IKM
with finite transfer matrix between the impurities using numerical
renormalization group.~\cite{Sakai} This discrepancy requires
further investigation.

\subsection{Linear conductance with potential scattering}
We generalize the linear conductance calculation of
Sec.~(\ref{se:Tne0}) for finite potential scattering. Using
Eqs.~(\ref{eq:flavorzx}) and (\ref{eq:flavormajorana}) the odd
current operator is
\begin{equation}
\label{eq:jo} j_o = 2 i \chi_2^{f} \left(\cos \phi_m \chi_1^{f}+\sin
\phi_m \chi_2^{X} {\rm{sgn}}(x)\right).
\end{equation}
The operator $a$ is now coupled to three free Majorana fields, and
its GF, Eq.~(\ref{eq:Ga}) generalizes to $\mathcal{G}_a(i \omega_n)
= \bigl( i \omega_n + i \lambda^2 {\rm{sgn}}(\omega_n)/2
\bigr)^{-1}$, where $\lambda^2$ is defined in Eq.~(\ref{eq:Tstar}).
Similarly,
\begin{equation}
-\langle \chi_i(x) \chi_j(y) \rangle_{i \omega_n} =
\mathcal{G}^{(0)}(i \omega_n,x,y) \delta_{ij}+h_{i} h_{j} \delta
\mathcal{G} (i \omega_n,x,y),  \nonumber
\end{equation}
where
\begin{equation}
\delta \mathcal{G} (i \omega_n,x,y) =  \mathcal{G}^{(0)}(i
\omega_n,x,0) \mathcal{G}_a(i \omega_n) \mathcal{G}^{(0)}(i
\omega_n,0,y).  \nonumber
\end{equation}

Generalizing Eq.~(\ref{eq:odd}) we obtain the odd current correlator
\begin{eqnarray}
\langle  j_o(x,\tau) j_o(y,0)
\rangle_{\lambda_1,\lambda_2,\lambda_3}
= \langle j_o(x,\tau) j_o(y,0) \rangle_{free} \nonumber \\
 +\bigl(\sin^2 \phi_m(\lambda_1^2+\lambda_3^2)-\cos^2 \phi_m(\lambda_2^2+\lambda_3^2)\bigr)\nonumber \\ \times 4 \mathcal{G}^{(0)} (i \omega_n,x,y)
 \delta
\mathcal{G} (i \omega_n,x,y).\nonumber
\end{eqnarray}
As a result the conductance has the scaling form
\begin{equation}
\label{eq:scaling} G/G_0 = 1-F\bigl[ T/T^* \bigr] \frac{\sin^2
\phi_m(\lambda_1^2+\lambda_3^2)-\cos^2
\phi_m(\lambda_2^2+\lambda_3^2)}{\lambda^2 \sin^2 \phi_m}.
\end{equation}
We see that the conductance at the free fixed point ($\lambda=0$) is
still given by $G_0= \frac{2 e^2}{h} \sin^2 \phi_m$. At $\lambda
\rightarrow \infty$ the Fermi liquid conductance is
\begin{equation}
\label{eq:GFL}G_{FL} =\frac{2e^2}{h} \frac{(\lambda_2)^2 +\cos^2
\phi_m (\lambda_3)^2 }{\lambda^2}.
\end{equation}
We may rewrite Eq.~(\ref{eq:scaling}) as
\begin{equation}
\frac{G-G_{FL}}{G_0 - G_{FL}} = 1 - F\bigl[T/T^* \bigr].\nonumber
\end{equation}

\subsection{Gan's theory and its relation to boundary Ising model}
\label{se:Gan} Gan presented a solution of the 2IKM, constructing an
effective Hamiltonian for a finite region in the phase diagram
around the critical point by controlled projection.~\cite{Gan95} The
effective Hamiltoanian is solved exactly not only at the critical
point but also for the surrounding Fermi-liquid phase. Excellent
agreement was found with numerical renormalization group and CFT, in
spite of the fact that the theory of Gan is not spin-SU(2)
invariant. We shall substantiate the relation of Gan's theory to the
CFT by showing explicitly that the operators at the critical point
have the same form for both theories. In the next section we will
use this approach to calculate the nonlinear conductance.

Gan theory uses the $SO(8)$ representation, and the two impurity
spins turn into a local fermion $d$, where $\{d,d^\dagger \} =1$.
Defining two Majorana fermions $a=\frac{d-d^\dagger}{\sqrt{2} i}$
and $b = \frac{d+d^\dagger}{\sqrt{2}}$, Gan's Hamiltonian in the P-H
symmetric case involves only the spin-flavor ($X$) sector, and can
be written as $H_{G}=H^{(0)}_{G}+ \delta H_{G}$ where
\begin{eqnarray}
\label{eq:Hgan} H^{(0)}_{G}&=&\frac{1}{2}\int dx \chi_2^X i
\partial_x
\chi_2^X,\nonumber \\
\delta H_{G}&=&2 i\sqrt{T_K} \chi_2^X(0) b-i(K-K_c)a b.
\end{eqnarray}
We shall show that for energy scales $\ll T_K$ this coincides with
the Ising model Eq.~(\ref{eq:HIsing}). To see this suppose $K=K_c$
and consider a mode expansion
\begin{eqnarray}
\nonumber \chi_2^X(x) = \sum^\Lambda_k (\varphi_k(x) \psi_k+
h.c.),~~~ b= \sum^\Lambda_k (u_k \psi_k+  h.c.),
\end{eqnarray}
where $\{ \psi_k , \psi^\dagger_{k`}\}=\delta(k-k')$, $\{ \psi_k ,
\psi_{k`}\}=0$, and where initially we choose $\Lambda \gg T_K$ as
an ultraviolet cutoff. In the basis of $\psi_k$ Gan's Hamiltonian is
equal to $H = \sum_k \epsilon_k \psi_k^\dagger \psi_k$. One can
obtain a Schr\"{o}dinger`s equation for the wave functions
$\varphi_k(x)$ and $u_k$ by equating the expansions of
$[H_G,\chi_2^X(x)]=[H,\chi_2^X(x)]$ and $[H_G,b]=[H,b]$. One obtains
\begin{eqnarray}
2 i \sqrt{T_K} \delta(x) u_k+i \partial_x \varphi_k(x) &=&
\epsilon_k
\varphi_k(x), \nonumber \\
-2 i \sqrt{T_K} \varphi_k(0) &=& \epsilon_k u_k. \nonumber
\end{eqnarray}
The solution is $\varphi_k(x) \propto e^{i k x}[ \theta(x)
\varphi^{(+)}_k+ \theta(-x) \varphi^{(-)}_k]$, $\varphi_k(0)
=\frac{1}{2}( \varphi^{(+)}_k+ \varphi^{(-)}_k)$, $u_k = \frac{2}{i
\epsilon_k}\sqrt{T_K}\varphi_k(0)$, $\varphi^{(-)}_k/\varphi^{(+)}_k
= e^{ 2 i \delta}$, $\tan \delta=\frac{2 T_K}{ \epsilon_k}$,
$\epsilon_k=-k$ (note that we work with left movers). While at
$T_K=0$ we have the BC $\chi_2^X(0^+) = \chi_2^X(0^-)$, we see from
the wave function that the effect of the first boundary term in $
H_G$ is to modify this BC to $\chi_2^X(0^+) =- \chi_2^X(0^-)$ for
energies $\ll T_K$. The key observation is that the following
operator identity holds if one restricts the mode expansion of its
LHS and RHS to energies below a cutoff $\Lambda \ll T_K$,
\begin{equation}
\label{eq:b} b  =\frac{1}{\sqrt{ T_K}} \chi_1(0),
\end{equation}
where $\chi_1(x) = \chi_2^X(x) {\rm{sgn}}(x)$. Physically this means
that at energy scales below $T_K$ the local operator $b$ is absorbed
into the field $\chi_2^X$ and changes its BC. Using the operator
identity Eq.~(\ref{eq:b}), we see that the term $\propto K-K_c$ in
$\delta  H_G$ is equivalent to the boundary operator in the Ising
model, Eq.~(\ref{eq:sigmaB}).

This establishes the connection between Gan's theory and the
boundary Ising model arising from the CFT solution, showing that
Gan's anisotropic theory describes correctly also the vicinity of
the isotropic fixed point.

\section{Crossover at finite bias}
\label{se:V} Gan's formulation of the QCP in the SO(8) Majorana
representation provides a direct way to calculate the nonlinear
conductance at finite source drain voltage along the crossover from
the NFL fixed point to the surrounding FL fixed points, including
the P-H symmetry breaking. Relegating the details of the calculation
based on the Keldysh technique to the appendix, our result is
\begin{eqnarray}
\label{eq:result} G &=& G_0 + G_S F[\frac{T}{T^*},\frac{eV}{T^*}]+
G_A
F'[\frac{T}{T^*},\frac{eV}{T^*}],\nonumber \\
F[t,v] &=& \frac{1}{4 \pi t} {\rm{Re}}~\psi_1~\left(
\frac{1}{2}+\frac{1}{4
\pi t}+\frac{i v}{2 \pi t}\right),\nonumber \\
F'[t,v] &=& \frac{1}{4 \pi t} {\rm{Im}}~\psi_1~\left(
\frac{1}{2}+\frac{1}{4 \pi t}+\frac{i v}{2 \pi t}\right),\nonumber
\\\frac{G_S}{\frac{2 e^2}{h}} &=& \frac{-\lambda_1^2 \sin^2 \phi_m+ \lambda_2^2
\cos^2 \phi_m+\lambda_3^2(1-2 \sin^2 \phi_m)}{\lambda^2},\nonumber
\\
\frac{G_A}{\frac{2 e^2}{h}} &=& \sin(2 \phi_m) \lambda_3
\frac{\lambda_2 \sin \phi_m+ \lambda_1 \cos \phi_m}{\lambda^2}.
\end{eqnarray}
Here $T^*,G_0,\phi_m$ and $\theta$ are given in Eqs.
(\ref{eq:Tstar}), (\ref{eq:G0}), (\ref{eq:phim1}) and
(\ref{eq:parametrization1}); $\psi_1(z)$ is defined below
Eq.~(\ref{eq:con}). This result is valid for $eV, T , T^* \ll T_K$.
When $T^* \gg T,eV$ the system is in the FL state and the nonlinear
conductance coincides with the linear conductance,
Eq.~(\ref{eq:GFL}), $G_{FL} = G_0+G_S$.

\begin{figure}[h]
\begin{center}
\includegraphics*[width=60mm]{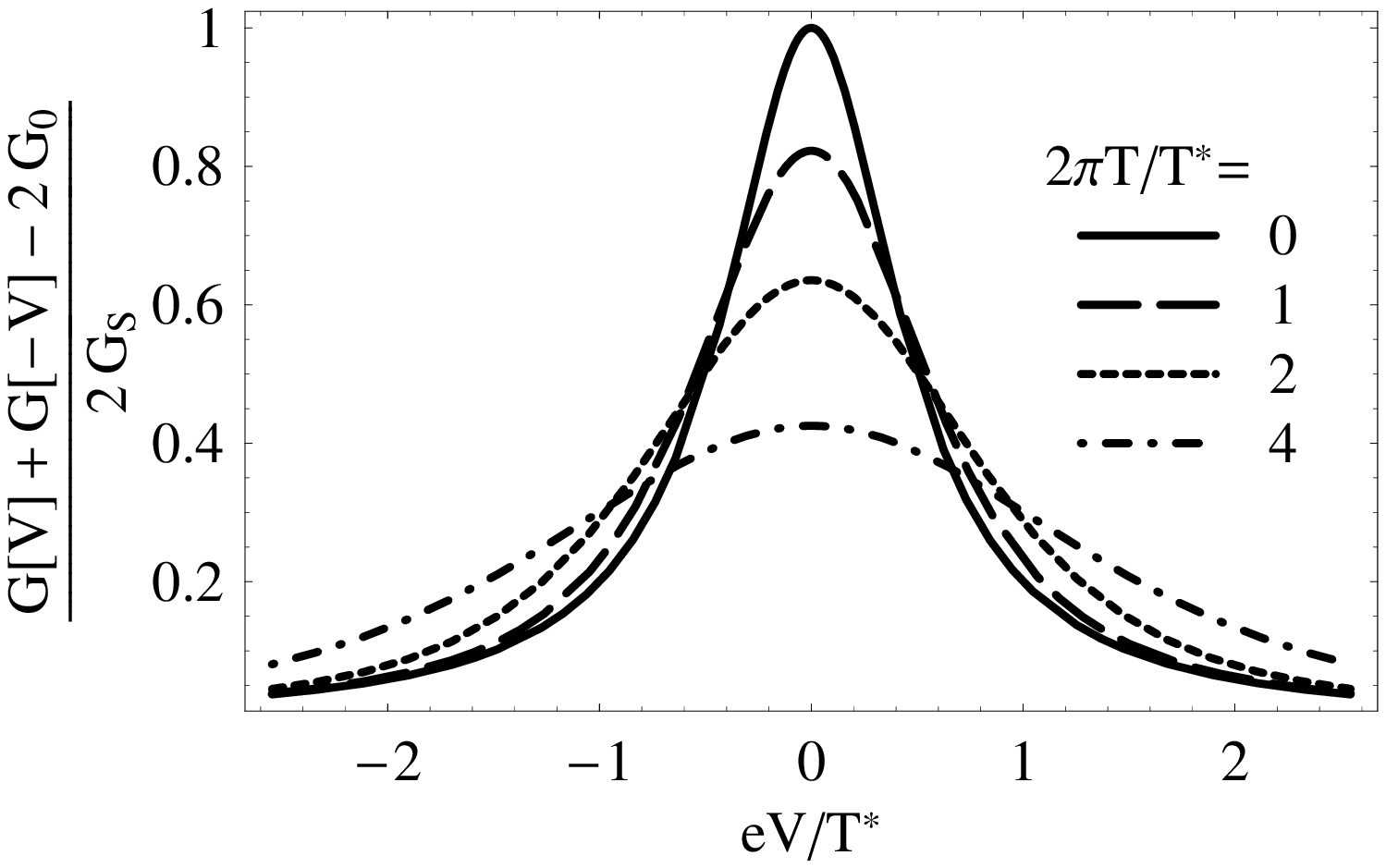}
\caption{Scaling function for the $V-$symmetric part of the
nonlinear conductance. \label{fg:Gsym}}
\end{center}
\end{figure}
\begin{figure}[h]
\begin{center}
\includegraphics*[width=60mm]{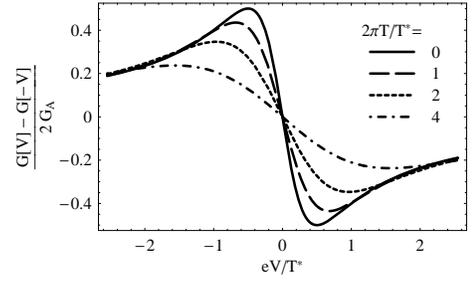}
\caption{Scaling function for the $V-$asymmetric part of the
nonlinear conductance. \label{fg:Gasym}}
\end{center}
\end{figure}

The scaling functions $F[t,v]$ and $F'[t,v]$ are symmetric and
asymmetric in $v$, respectively; see Figs.~(\ref{fg:Gsym}) and
(\ref{fg:Gasym}). The asymmetric component is a new feature in the
parallel QD, as compared to series QD where $\phi_m=0$. Having $G[V]
\ne G[-V]$ is a signature of interactions, since the Landauer
noninteracting formula\cite{Landauer57} leads to $G[V] =G[-V]$. This
leads to a universal rectification effect. This rectification effect
is odd under parity, $\lambda_3 \to - \lambda_3$. Note however that
it does not have a well defined transformation property with respect
to $\Phi \to - \Phi$. To check the symmetry properties of our
results we considered the two impurity Anderson model for our
 model [Fig.(\ref{fg:1})] to first order in the (intra-dot) interaction $U$. While at
$U=0$ we have $G[V] =G[-V]$, which follows from Landauer formula, to
first order in $U$ we get a finite $G[V] - G[-V]$. This asymmetric
behavior of the conductance follows from an asymmetric dependence of
the occupation of the dots on voltage. This simple limit gives the
same symmetry properties of $G[V] - G[-V]$ compared to the QCP,
namely the rectification effect is odd under a parity, and does not
have a well defined symmetry property with respect to $\Phi \to
-\Phi$.

At energy scales comparable to $T_K$ the conductance has additional
voltage and temperature dependence due to irrelevant operators at
the QCP. The leading irrelevant operator is $H_{irr}=T_K^{-1/2} i
\partial_x \chi_1(x) a|_{x=0}$, with dimension $\Delta = 3/2$.\cite{Affleck92,Affleck95} In the
proposed realization of the 2IKM of Zar\'{a}nd \emph{et.
al.}~\cite{Zarand06} it leads to the conductance correction $\delta
G \propto \sqrt{\frac{T}{T_K}}$, characteristic of a NFL fixed
point. However in the present system the irrelevant operator gives a
nonzero correction only to fourth order, leading to $\delta G
\propto \left({\frac{T}{T_K}} \right)^2$, as we outline below. The
$a$-GF has an additional self energy $\Sigma ^R=-i \omega^2/T_K$,
\begin{equation}
G_a^R (\omega)= \frac{1}{\omega+i T^*/2} \to \frac{1}{\omega+i
(T^*/2 + \omega^2/T_K)}.\nonumber
\end{equation}
This has poles at $\omega= -i T_K \frac{1}{2} (1 \pm \sqrt{1-\frac{2
T^*}{T_K}})$. For $T^* \ll T_K$ we have
\begin{equation}
G_a^R (\omega) \cong \frac{1}{\omega+i T^*/2} - \frac{1}{\omega+i
T_K}.\nonumber
\end{equation}
Qualitatively, the irrelevant correction at finite $T_K$ has the
same form of the fixed point conductance with $T^*/2 \to T_K$.
Indeed at energy scales smaller than $T^*$, the latter has quadratic
dependence on $T/T^*$ and $eV/T^*$.\cite{Sela08} It should be
pointed out that to fourth order in $H_{irr}$ it is no longer
consistent to disregard more irrelevant operators of dimension
$\Delta = 2$. However their inclusion leads only to the modification
of the effective Kondo temperature in the corrections $(T/T_K)^2$
and $(eV/T_K)^2$.

\begin{figure}[h]
\begin{center}
\includegraphics*[width=70mm]{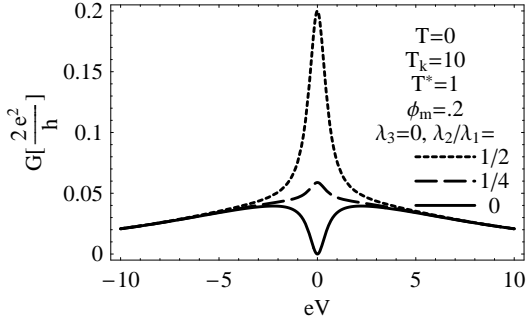}
\caption{Nonlinear conductance at $\lambda_3=0$ for
$\lambda_2/\lambda_1=0,1/4,1/2$. The line shape consists of a narrow
peak or dip structure of width $T^*$, superimposed on top of a wide
peak of width $T_K$. \label{fg:l3equal0A}}
\end{center}
\end{figure}

\begin{figure}[h]
\begin{center}
\includegraphics*[width=70mm]{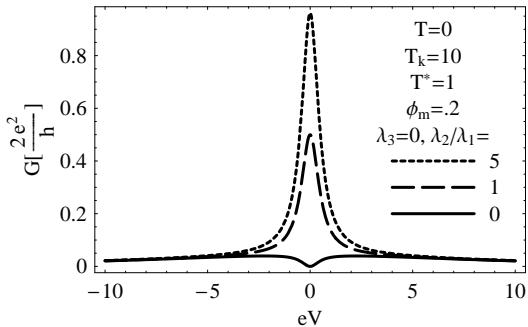}
\caption{Nonlinear conductance at $\lambda_3=0$ for
$\lambda_2/\lambda_1=0,1,5$, reaching the unitary limit when the
relevant perturbation is dominated by potential scattering, namely
$\lambda_2 \gg \lambda_1$. \label{fg:l3equal0}}
\end{center}
\end{figure}

In Figs.~(\ref{fg:l3equal0A}) and (\ref{fg:l3equal0}) the
conductance is plotted in the parity-symmetric case at $\lambda_3=0$
and zero temperature as function of source drain voltage, for
different ratios $\lambda_2/\lambda_1$. The generic behavior of
$G[V]$ consists of a wide peak of width $T_K$ and height $G_0$, with
a superimposed narrow structure (peak or dip) of width $T^*$, with
height $G_S$ (relative to the background $G_0$). Note that $G_S$ is
positive(negative) for $\lambda_1^2 \sin^2 \phi_m <(>) \lambda_2^2
\cos^2 \phi_m+\lambda_3^2(1-2 \sin^2 \phi_m)$, leading to a narrow
peak(dip). When $\lambda_3=0$ and $\lambda_1 \tan \phi_m \ll
\lambda_2$, Eq.~(\ref{eq:result}) predicts a peak amplitude close to
the unitary limit $2e^2/h$. For this case, we mention that when
$T^*_{LR}$ and $|K-K_c|\gtrsim T_K$, our results do not apply, and
we expect a splitting of this peak as a function of
$V$.~\cite{Jeong01,moreddot} We can obtain this behavior on a
qualitative level by going back to the high energy $E \sim T_K$
description with Eq.~(\ref{eq:Hgan}).

\begin{figure}[h]
\begin{center}
\includegraphics*[width=70mm]{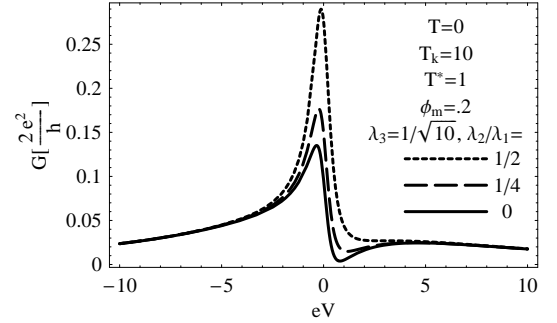}
\caption{Nonlinear conductance at finite $\lambda_3=1/\sqrt{10}$ for
$\lambda_2/\lambda_1=0,1/4,1/2$, showing asymmetric features.
\label{fg:l3NOTequal0A}}
\end{center}
\end{figure}

\begin{figure}[h]
\begin{center}
\includegraphics*[width=70mm]{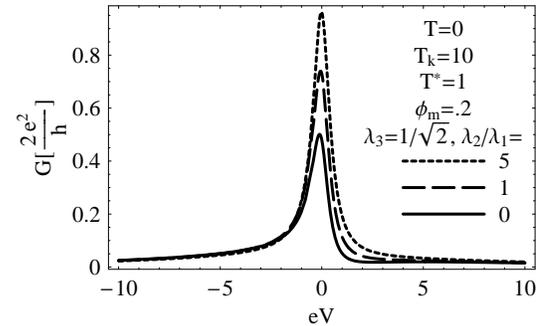}
\caption{Nonlinear conductance at finite $\lambda_3=1/\sqrt{2}$  for
$\lambda_2/\lambda_1=0,1,5$. \label{fg:l3NOTequal0}}
\end{center}
\end{figure}

In Figs.~(\ref{fg:l3NOTequal0A}) and (\ref{fg:l3NOTequal0}) we plot
the conductance under the same conditions except
$\lambda_3=1/\sqrt{10}$ and $\lambda_3=1/\sqrt{2}$, respectively,
showing asymmetric behavior. When $G_A>0(G_A<0)$ [defined in
Eq.~(\ref{eq:result})], the slope of the conductance at $V=0$,
$\frac{dG}{dV}|_{V=0}$, is negative(positive). The sign of $G_A$ is
changed under a parity transformation ($\lambda_3 \to - \lambda_3$),
but it also depends on the sign of the combination $\left(\sin
(2\phi_m) (\lambda_2 \sin \phi_m+ \lambda_1 \cos \phi_m)\right)$.

\subsection{Observability} \label{se:observability} In this subsection
we discuss the realizability of the critical point in real
experiment. Dealing with a repulsive critical point, the first
condition we are concerned with is the smallness of the relevant
perturbations, $T^* \ll T_K$. Secondly, we shall list some marginal
corrections.

In order to tune $K = K_c > 0$, it is needed to reduce the
ferromagnetic contribution $K_{RKKY}$ [Eq.~(\ref{eq:RKKY})] compared
to $K_{12}$ [Eq.~(\ref{eq:K12})], either (i) by setting $\Phi=\pi$,
which is sufficient in the ideal situation where the device is
perfectly parity symmetric, or (ii) in the more generic and
realistic case, where parity symmetry is only approximate, by
creating \emph{large asymmetry} \begin{equation}
\label{eq:asymmetry} t_2 / t_1 \ll \frac{t_{12}}{U (\nu J)}\sim
\sqrt{\frac{T_K}{U}} \frac{1}{\nu J}.
\end{equation}
The limit $t_2=0$ corresponds to the the series QD. In either case,
using Eq.~(\ref{eq:K12}), the condition $K = K_c \sim T_K$ is
achieved by tuning the inter-dot coupling $t_{12} \sim \sqrt{U
T_K}$.

At $K=K_c$ ($\lambda_1=0$) Eqs.~(\ref{eq:Tstar}) and
(\ref{eq:lambdas}) give the crossover scale
\begin{equation}
 T^*_{LR}=T^*|_{K=K_c} =T_K \left(  ({\rm{Re}}\nu V_L^R)^2+(\nu V_B)^2
\right),\nonumber
\end{equation}
where $V_B$ and $V^R_L $ are given in Eqs.~(\ref{eq:VAB}) and
(\ref{eq:V}).

In the parity symmetric case (i) we have $V_B=0$, and $V_L^R$ is
real and dominated by its second term in Eq.~(\ref{eq:V}) because
$\Phi = \pi$, leading to
\begin{equation}
\label{eq:scale} T^*_{LR} /T_K \sim |\nu V_{L}^R|^2 \sim (\nu J)^2
\frac{T_K}{U} \ll 1,
\end{equation}
 as required for the validity of the critical theory.

In the more realistic case (ii), on top of Eq.~(\ref{eq:asymmetry})
we bound $t_2/t_1$ from below,
\begin{equation}
\sqrt{\frac{T_K}{U}} \sim t_{12}/U \lesssim t_2/t_1 \ll
\sqrt{\frac{T_K}{U}}, \nonumber
\end{equation} such that $V_L^R$ is
dominated by the first term in Eq.~(\ref{eq:V}). In addition we
demand approximate parity symmetry,
\begin{equation}
\frac{|t_{1L}-t_{2L}|}{t_1} \ll \sqrt{T_K/U} (t_1/t_2),~~~
\sin(\Phi_L - \Phi_R) \ll 1,\nonumber
\end{equation}
such that $|V_B| \ll {\rm{Re}} V_{L}^R$. Here $t_1 =
\frac{t_{1L}+t_{2L}}{2}$ and $t_2 = \frac{t_{2L}+t_{1L}}{2}$. It
leads to $ T^*_{LR} /T_K \sim |\nu V_{LR}|^2 \sim (\nu J)^2 \left(
\frac{t_2}{t_1}\right)^2 \ll 1$, as required for the validity of the
critical theory.

Next we estimate the marginal corrections. Spin $SU(2)$ symmetry is
broken by the Zeeman energy $E_Z=g \mu_B B(S_1^z+S_2^z)$. This leads
to a marginal operator\cite{Gan95} which  reads in the Bose-Ising
representation $\vec{\phi}~ \epsilon$. In GaAs QDs, the Zeeman
energy is reduced due to a small g-factor: for the experimental
conditions in Ref.~[\onlinecite{Gordon98}] $T_K$ corresponds to a
magnetic field of few tesla, or equivalently to $\sim 10^3$ flux
quanta in a area of $\mu m^2$; for a magnetic field corresponding to
$\Phi= \pi$ we have $\langle E_Z \rangle \sim 10^{-3}T_K $, leading
to small marginal correction to the conductance.

Other marginal operators allowed at the QCP are the inter- and
intra- channel PS, Eqs.~(\ref{eq:VLR}) and (\ref{eq:VLL}). Those are
expected to introduce small corrections of $O((\nu J)^2)$ the
conductance. Part of those operators break parity symmetry. So, the
parity symmetry Eq.~(\ref{eq:parity}) is not required to hold
exactly. Indeed the QCP has been observed numerically for a broken
parity Hamiltonian in Ref.~[\onlinecite{Zarand06}].

\section{Conclusions}
\label{se:conclusion} We studied double quantum dots in the vicinity
of the quantum critical point of the 2-impurity Kondo model. In the
P-H symmetric model we used a mapping to the boundary Ising model
with finite boundary magnetic field, to calculate the finite
temperature crossover of the conductance from the QCP to the stable
fixed points. This method generalizes the CFT approach, which
addresses only the vicinity of the fixed points. We used this method
to relate the conductance of the proposed system of Zar\'{a}nd
\emph{et. al.}~\cite{Zarand06} to the one-point function of the
magnetization operator in the boundary Ising model which can be
calculated numerically.

Using the method developed by Gan, we solved the general and
experimentally relevant case with potential scattering, and found
the nonlinear conductance at finite temperature along the
multidimensional crossover from QCP to surrounding FL states.
Compared to the series double QD, we found that in the general
configuration the universal scaling function contains both symmetric
and asymmetric terms in the source drain voltage, leading to a
current rectification.

\section{Acknowledgments}
We thank A. Aharony, D. Eigler, O. Entin-Wohlman, J. Folk, D.
Goldhaber-Gordon, S. Loth and J. Malecki for very helpful
discussions. This work was supported by NSERC (ES $\&$ IA) and CIfAR
(IA).

\appendix
\section{Calculation of the nonlinear conductance using Keldysh Green function technique}
 \label{se:Keldysh} We briefly recall basic concepts of the nonequilibrium
formulation. Then the problem at hand will be addressed, and the
calculation of the  nonlinear conductance will be outlined.

One usually assumes that the system is in equilibrium at some
initial time, taken here to be $t=-\infty$. A perturbation $H_1$ is
turned on adiabatically in time, $H=H_0+e^{\eta t} H_1$ to drive the
system out of equilibrium. The expectation value of an operator such
as the current $I$ is given by its trace in the Heisenberg picture
at $t=0$ weighted by the initial distribution function,
\begin{equation}
\langle I \rangle ={ \rm{Tr}} \{  e^{-\beta H_0}
u^\dagger(0,-\infty) \hat{I} u(0,-\infty) \},\nonumber
\end{equation}
where $u(t_0,t)=\mathcal{T} \exp \bigl(-i \int_{t_0}^t dt' H(t')
\bigr)$ and $\mathcal{T}$ is the time ordering operator. In order to
employ Wick's theorem, one transforms to the interaction picture,
$\langle I \rangle ={ \rm{Tr}} \{ e^{-\beta H_0}
u_I^\dagger(0,-\infty) \hat{I}_I u_I(0,-\infty) \}$, where
$u_I(t_0,t)=\mathcal{T} \exp \bigl(-i \int_{t_0}^t dt' (H_{1})_I(t')
\bigr)$, and $\mathcal{O}_I(t) = e^{i H_0 t} \mathcal{O} e^{-i H_0
t}$. Following Keldysh, for a perturbative expansion of this
quantity it is convenient to introduce 4 types of
GFs,~\cite{Keldysh65}
\begin{eqnarray}
G^{11}(1,1') &=& -i \langle \mathcal{T} \chi(1)\chi(1')
\rangle,\nonumber \\
G^{12}(1,1')&=&G^<(1,1') = i \langle
\chi(1') \chi(1) \rangle, \nonumber \\
G^{21}(1,1')&=&G^>(1,1') = -i \langle \chi(1)\chi(1')
\rangle,\nonumber \\
G^{11}(1,1') &=& -i \langle \tilde{\mathcal{T}}\chi(1)\chi(1')
\rangle. \nonumber\end{eqnarray} Here $\tilde{\mathcal{T}}$ is the
anti-time ordering operator. It is convenient to
consider an alternative set of GFs, by defining the Keldysh GF matrix $\underline{G} =\left(%
\begin{array}{cc} G^R
 & G^<  \\
0
 &  G^A\\
\end{array}%
\right)$, where
\begin{eqnarray}
G^{R,A}(1,1') &=& \mp i \theta[\pm(t_1 - t_{1}')]\langle \{
\chi(1),\chi(1') \}_+ \rangle.\nonumber \end{eqnarray}
Given a self energy, $\underline{\Sigma} =\left(%
\begin{array}{cc} \Sigma^R
 & \Sigma^<  \\
0
 &  \Sigma^A\\
\end{array}%
\right)$, the Keldysh GF matrix has the expansion
\begin{equation}
\label{eq:series} \underline{G}(\omega) =
\underline{G}^{(0)}(\omega)+ \underline{G}^{(0)}(\omega)
\underline{\Sigma}(\omega) \underline{G}^{(0)}(\omega)+ ...,
\end{equation}
where matrix multiplication in Keldysh space is understood and
$A(\omega) = \int dt e^{i \omega t} A(t)$. This leads to the Dyson
equation for the retarded/advanced components of $\underline{G}$,
\begin{equation}
\label{eq:Dyson} G^R(\omega) = {G^R}^{(0)}(\omega)+
{G^R}^{(0)}(\omega) \Sigma^R(\omega) G^R(\omega),
\end{equation}
and to the Keldysh equation
\begin{equation}
\label{eq:Keldysh} G^<=G^R \Sigma^< G^A+(1+G^R \Sigma^R)
{{G^<}^{(0)}} (1+\Sigma^A G^A).
\end{equation}

We now apply this scheme to our problem with
\begin{eqnarray}
H_0 &=&\sum_{j=1}^2 \sum_{A=c,s,f,X}\frac{1}{2}\int_{-\infty}^\infty
dx
 \chi_j^A i
\partial_x \chi_j^A\nonumber \\
&+& \frac{eV}{2}\sum_\alpha(N_{L \alpha} - N_{R \alpha}),\nonumber \\
H_1 &=& \delta H_G+i \lambda_2 \chi_1^f a+i \lambda_3 \chi_2^f a,
\nonumber
\end{eqnarray}
where $N_{i \alpha}=\int dx {\psi^\dagger}^{i \alpha} \psi_{i
\alpha}$, ($i=L,R$). Here $H_0$ is the $J=0$ fixed point
Hamiltonian, including the source drain voltage $V$, and $\delta
H_G$ is given in Eq.~(\ref{eq:Hgan}). It is more convenient to use
$\delta H_G$, which includes the local $b$ operator, rather than the
first term in $\delta H$, $i\lambda_1 \chi_2^X(x) {\rm{sgn}}(x) a$.
Both formulations should give the same result for energy scales $\ll
T_K$, as we showed generally in Sec.~\ref{se:Gan}. At $t=- \infty$
the system consists of two decoupled leads at equilibrium with
different chemical potentials. It is convenient to make a change of
basis, in which the operator $Y = \frac{1}{2}\sum_\alpha(N_{L
\alpha} - N_{R \alpha})= \frac{1}{2}\int_{-\infty}^\infty dx j_o(x)$
is diagonal. Using Eq.~(\ref{eq:jo}) for $j_o$, we see that
\begin{equation}
Y = \int_{-\infty}^\infty dx \alpha^\dagger \alpha = i
\int_{-\infty}^\infty dx \alpha_- ~ \alpha_+,\nonumber
\end{equation}
where we defined new fermions $\alpha$ and $\beta$: $\alpha =
\frac{\alpha_+ - i \alpha_-}{\sqrt{2}}$, $\beta = \frac{\beta_+ - i
\beta_-}{\sqrt{2}}$, in terms of the 4 Majorana fermions
$\alpha_\pm$ and $\beta_{\pm}$ given by
\begin{eqnarray}
\alpha_+&=&\chi_2^f,~~~\beta_+ = \chi_1^X,  \nonumber \\
\alpha_-&=&-(\cos \phi_m \chi_1^f+\sin \phi_m \chi_2^X),
\nonumber \\
\beta_-&=&(\sin \phi_m \chi_1^f-\cos \phi_m \chi_2^X).\nonumber
\end{eqnarray}
We see that the voltage raises the chemical potential of the
$\alpha$ fermions by $eV$, whereas the chemical potential for the
$\beta$ fermions remains zero. The system is at equilibrium at $t =
- \infty$ since in this case the $\alpha$ and $\beta$ fermion
numbers are conserved. At $t > - \infty$, $H_1$ leads to the current
operator $\hat{I} = i [Y,H_1]$ which drives the system out of
equilibrium and is given by
\begin{eqnarray}
\label{eq:I} \hat{I} = -2 i \sqrt{T_K} \sin \phi_m \alpha_+(0) b
\nonumber \\
-i \lambda_2 \cos \phi_m \alpha_+(0) a-i \lambda_3
\alpha_-(0) a.
\end{eqnarray}
We shall express the expectation value of the current by the Green
functions $\underline{G}_{ \nu \mu}(t)$ where the indices refer to
the fermion local operators $\nu=(a,b)=(1,2)$ and
$\mu=(\alpha_+(x=0),
\alpha_-(x=0),\beta_+(x=0),\beta_-(x=0))=(1,2,3,4)$. Using
Eq.~(\ref{eq:I}) the current expectation value reads
\begin{eqnarray}
\label{eq:currentG} \langle I(t=0) \rangle =-2 \sqrt{T_K} \sin
\phi_m G^<_{b \alpha_+}(t=0)\nonumber \\ - \lambda_2 \cos \phi_m
G^<_{a \alpha_+}(t=0)-\lambda_3 G^<_{ a \alpha_-}(t=0).
\end{eqnarray}

We construct the exact GFs [appearing in Eq.~(\ref{eq:currentG})]
from the free GFs calculated at $t= - \infty$ ($H_1 =0$): for
$\mu,\mu'=(\alpha_+, \alpha_-,\beta_+,\beta_-)=(1,2,3,4)$, one finds
\begin{eqnarray}
\label{eq:Gfree} (G^{R,A})^{(0)}_{\mu \mu'} &=& \frac{\mp i}{2}
\delta_{\mu \mu'},\nonumber \\(G^{<})^{(0)}_{\alpha_\pm,\alpha_\pm}
&=&i \tilde{f}(\omega),~~~ (G^{<})^{(0)}_{\beta_\pm,\beta_\pm} = i
f(\omega),\nonumber
\\(G^{<})^{(0)}_{\alpha_\pm, \alpha_\mp}&=&\pm \frac{1
}{2}[f(\omega-eV)-f(\omega+eV)] .
\end{eqnarray}
Here $f(x)=(1+e^{x/T})^{-1}$, $\tilde{f}(x)=\frac{
1}{2}[f(x+eV)+f(x-eV)] $. Note that the voltage couples the two
Majorana fermions $\alpha_+$ and $\alpha_-$, and here we assumed a
band width $ \gg \omega, V, T$. The free GF for the local Majorana
fermions $\nu=(a,b)=(1,2)$ is $(G^R)^{(0)}_{\nu \nu'} = \delta_{ \nu
\nu'} (\omega+i \delta)^{-1}$, where $\delta$ is a positive
infinitesimal. We write $H_1$ in a convenient form
\begin{equation} H_1 =-i \sqrt{T_K} \lambda_1 a b - i \sum_{\mu=1}^4
\sum_{\nu=1}^2 (\alpha_+,\alpha_-,\beta_+,\beta_-)_\mu  \Lambda_{\mu
\nu} (a,b)_{\nu},\nonumber
\end{equation}
where \begin{eqnarray}\label{eq:bigLambda}
(\Lambda_{11},\Lambda_{21},\Lambda_{31},\Lambda_{41})&=&(-\lambda_3,\lambda_2
\cos \phi_m,0,-\lambda_2 \sin \phi_m), \nonumber \\
({\Lambda_{12}},\Lambda_{22},\Lambda_{32},\Lambda_{42})&=&2
\sqrt{T_K}(0, \sin \phi_m,0, \cos \phi_m).
\end{eqnarray}

We obtain the full GF $\underline{G}_{ \nu  \nu'} $ for
$\nu,\nu'=a,b$ as follows. First suppose $K=K_c$ ($\lambda_1=0$); we
denote the different GFs and self energies in this case as
$\bar{G}$, $\bar{\Sigma}$, respectively. At $K=K_c$ the self energy
matrix is
\begin{equation}
\underline{\bar{\Sigma}}_{\nu \nu'} =- \Lambda_{\mu \nu}
\underline{G}^{(0)}_{\mu \mu'} \Lambda_{\mu' \nu'},~~~
({\rm{repeated~indices~summed}}).\nonumber \end{equation}
Eqs.~(\ref{eq:Gfree}) and (\ref{eq:bigLambda}) give
$\bar{\Sigma}^R_{a a} = -\frac{i}{2}(\lambda_2^2+\lambda_3^2)$,
$\bar{\Sigma}^R_{bb}= -2 i T_K$, $\bar{\Sigma}^R_{ab} =
\Sigma^R_{ba}=0$. Eq.~(\ref{eq:Dyson}) gives $\bar{G}^R_{aa} =
(\omega+i \frac{\lambda_2^2+\lambda_3^2}{2})^{-1}$, $\bar{G}^R_{bb}
= (\omega+2 i T_K)^{-1}$, $\bar{G}^R_{ab}=\bar{G}^R_{ba} =0$. For
energies $\ll T_K$ we can approximate $\bar{G}^R_{bb} = (2 i
T_K)^{-1}$. For the lesser GF Eq.~(\ref{eq:Keldysh}) gives
\begin{equation}
\label{eq:Gbarlesser} \bar{G}^<=\bar{G}^R \bar{\Sigma}^< \bar{G}^A,
\end{equation}
where matrix equation and multiplication in $ab$ space is
understood.

For $K \ne K_c$ the full matrix GF $\underline{G}_{\nu \nu'} $ can
be calculated from the series Eq.~(\ref{eq:series}) where
$G^{{(0)}}\to \bar{G}$ and
\begin{equation}
\Sigma^R=\Sigma^A = \sqrt{T_K} \lambda_1 \tau^y,\nonumber
\end{equation}
$\Sigma^<=0$, where $\tau^y$ acts in $ab$ space.
Eq.~(\ref{eq:Dyson}) gives
\begin{eqnarray}
G_{aa}^R(\omega)&=&(\omega+i \lambda^2/2)^{-1}, \nonumber \\
G_{bb}^R (\omega)&=& (2 i T_K)^{-1}-(\lambda_1^2/4 T_K)(\omega+i
\lambda^2/2)^{-1}, \nonumber \\
G_{ab(ba)}^R(\omega)&=&\mp (\lambda/2 \sqrt{T_K}) (\omega+i
\lambda^2/2)^{-1}.\nonumber \end{eqnarray} For $G^<$, since
$\Sigma^<=0$ we are left with the second term of
Eq.~(\ref{eq:Keldysh}), which simplifies to [using
Eq.~(\ref{eq:Gbarlesser}) and (\ref{eq:Dyson})]
\begin{equation} G^<=G^R
\bar{\Sigma}^< G^A,\nonumber \\
\end{equation}
where matrix equation and multiplication in $ab$ space is
understood.

The GFs appearing in the current Eq.~(\ref{eq:currentG}) satisfy the
Dyson equation \begin{equation} \underline{G}_{\nu \mu}(t=0)=\int
\frac{d\omega}{2 \pi} \sum_{\nu',\mu'}\underline{G}_{\nu
\nu'}(\omega) i \Lambda_{\mu' \nu'} \underline{G}^{(0)}_{\mu'
\mu}(\omega). \end{equation} To evaluate $\underline{G}^<_{\nu \mu}$
we use the identity $(\underline{A}~ \underline{B})^< =
\underline{A}^< \underline{B}^A+\underline{A}^R \underline{B}^<$.

We encounter two types of integrals for the current:
\begin{eqnarray}
I'[V,T,\lambda^2]=i \int d\omega
\frac{f(\omega-eV)-f(\omega+eV)}{\omega+ i \lambda^2/2}\nonumber \\
={\rm{Im}}~ \psi\left(\frac{1}{2}+\frac{\lambda^2/2+i e V}{2 \pi T} \right) ,\nonumber \\
I''[V,T,\lambda^2]=\int d\omega
\frac{f(\omega)-\tilde{f}({\omega})}{\omega+ i \lambda^2/2}\nonumber
\\
=\psi\left(\frac{1}{2}+\frac{\lambda^2/2}{2 \pi T}
\right)-{\rm{Re}} ~\psi\left(\frac{1}{2}+\frac{\lambda^2/2+i e V}{2
\pi T} \right).\nonumber
\end{eqnarray}
Note that $I'[V] =- I'[-V]$, and $I''[V] = I''[-V]$. From these
results one can readily obtain the result for the nonlinear
conductance, Eq.~(\ref{eq:result}).

\end{document}